**Motor antagonism dictates emergent dynamics in active double networks tuned by crosslinkers**

*Ryan J. McGorty[1], Christopher J. Currie[1], Jonathan Michel[2], Mehrzad Sasanpour[1], Christopher Gunter[3], K. Alice Lindsay[4], Michael J. Rust[5], Parag Katira[3], Moumita Das[2], Jennifer L. Ross[4], Rae M. Robertson-Anderson[1,*]*

[1]Department of Physics and Biophysics, University of San Diego, San Diego, California 92110, USA
[2]School of Physics and Astronomy, Rochester Institute of Technology, Rochester, New York 14623, USA
[3]Department of Mechanical Engineering, San Diego State University, San Diego, CA 92182, USA
[4]Department of Physics, Syracuse University, Syracuse, New York 13244, USA
[5]Department of Molecular Genetics and Cell Biology, University of Chicago, Chicago, Illinois 60637, USA
*randerson@sandiego.edu

**Abstract**

The cellular cytoskeleton relies on diverse populations of motors, filaments, and binding proteins acting in concert to enable non-equilibrium processes ranging from mitosis to chemotaxis. The cytoskeleton's versatile reconfigurability, programmed by interactions between its constituents, make it a foundational active matter platform. However, current active matter endeavors are limited largely to single force-generating components acting on a single substrate–far from the composite cytoskeleton in cells. Here, we engineer actin-microtubule composites, driven by kinesin and myosin motors and tuned by crosslinkers, to ballistically restructure and flow with speeds that span three orders of magnitude depending on the composite formulation and time relative to the onset of motor activity. Differential dynamic microscopy analyses reveal that kinesin and myosin compete to delay the onset of acceleration and suppress discrete restructuring events, while passive crosslinking of either actin or microtubules has an opposite effect. Our minimal advection-diffusion model and spatial correlation analyses correlate these dynamics to structure, with motor antagonism suppressing reconfiguration and de-mixing, while crosslinking enhances clustering. Despite the rich formulation space and emergent formulation-dependent structures, the non-equilibrium dynamics across all composites and timescales can be organized into three classes–slow isotropic reorientation, fast directional flow, and multimode restructuring. Moreover, our mathematical model demonstrates that diverse structural motifs can arise simply from the interplay between motor-driven advection and frictional drag. These general features of our platform facilitate applicability to other active matter systems, and shed light on diverse ways that cytoskeletal components can cooperate or compete to enable wide-ranging cellular processes.

**Significance**

The cell cytoskeleton is a composite of protein filaments and crosslinkers pushed out-of-equilibrium by molecular motors to mediate wide-ranging processes from migration to morphogenesis. The cytoskeleton is, thus, paradigmatic active matter and its composite nature is one of its hallmarks. Yet, state-of-the-art active matter focuses on single force-generating components and substrates. Here, we engineer cytoskeleton composites driven by dual motors to contract, flow, and restructure into diverse morphologies from interpenetrating scaffolds to phase-separated clusters. Competition between the activities of the two motors delays restructuring and suppresses de-mixing, while filament crosslinking has an opposite effect. Our bio-inspired active composites, bringing reconstituted systems closer to mimicking cytoskeletal complexity, are foundational for diverse materials applications from wound-healing to soft-robotics.

## Introduction

The cytoskeleton is a dynamic, non-equilibrium material comprising protein filaments, including actin, microtubules and intermediate filaments, as well as motor proteins, such as myosins and kinesins, that actively push and pull on the protein filaments (1–8). Crosslinking proteins also connect and bundle filaments as needed for cellular processes (9–12). This complex composite continuously restructures and reconfigures in response to demands of the cell, to enable diverse processes from cytokinesis to mechano-sensing (3–5,7,8,13–21). In vitro systems of reconstituted cytoskeletal proteins, which display rich and tunable dynamics, are also intensely studied as model active matter platforms to shed light on the non-equilibrium physics underlying force-generating, reconfigurable systems (7,12,19,22–40).

Interacting networks of semiflexible actin filaments and rigid microtubules provide tensile and compressive strength to the cytoskeleton while allowing for cell mobility, key to processes such as division and chemotaxis (15,16,41–45). Further, recent studies have shown that in vitro actin-microtubule composites exhibit emergent mechanical properties that are not a simple sum of the single component systems (46–48). For example, composites with comparable concentrations of actin and microtubules display both enhanced filament mobility and increased stiffness (46), as well as an emergent non-monotonic dependence of elasticity on actin crosslinking (47).

More recently, myosin II minifilaments have been incorporated into actin-microtubule composites, showing that synergistic interactions between actin and microtubules prevent disordered flow and rupturing often seen in actomyosin networks without crosslinkers (26–28). These studies have also shown that composites can be tuned to display enhanced mechanical strength (27), coordinated motion of actin and microtubules, sustained ballistic contraction, and mesoscale restructuring (26, 28)–all in the absence of crosslinking proteins to chemically connect the filaments.

Microtubule-based active matter systems have also been engineered using clusters of kinesin motors that crosslink and pull on microtubule bundles to create active nematics (23,24,30,31,34,35,49–55). In these systems, kinesins generate long-lasting turbulent flows by cyclically extending, buckling, fracturing, and healing bundles (49). More recently, actin has been incorporated into active MT fluids, resulting in turbulent flow, contraction, or formation of layered asters (29).

The distinct dynamics and structures that kinesin-driven and myosin-driven systems display begs the question as to how different active components and substrates cooperate or compete to control cellular processes. While composite active matter is beginning to be developed to introduce more control and tunability over single-substrate systems (26–29,56), incorporating two active components that act on distinct substrates represents a paradigm shift in active matter. Beyond the cellular relevance, such designs can elucidate general principles for non-equilibrium programmable materials that can reconfigure and generate force; and determine how to enhance programmability and expand the dynamical and structural phase space by altering the active and static nature of crosslinkers and the substrates on which they act.

Here, we engineer actin-microtubule composites that undergo a rich combination of advective flow, contraction, and multi-mode restructuring driven by kinesin *and* myosin motors. These dynamics are coupled to distinct time-evolving structures that range from interpenetrating actin-microtubule scaffolds to microscale phase-separated amorphous clusters. We couple differential dynamic microscopy (DDM) with particle image velocimetry (PIV) to discover that competition between kinesin-microtubule activity and actomyosin activity delays the onset of rapid restructuring while crosslinking of either actin or microtubules accelerates the time-evolution of active dynamics. Our advection-diffusion model and spatial correlation analyses correlate the dynamic antagonism that we observe with suppressed de-mixing of double-motor composites, and the crosslinker-mediated acceleration with enhanced restructuring and clustering. Despite these complexities, we find that the broad phase space of active dynamics can be organized into three general Classes with distinct types and rates of ballistic motion.

**Results and Discussion**

*Active cytoskeleton composite design and formulation-structure phase space.* We first describe our design strategy for realizing an active matter system that has two force-generating components that act on two distinct, co-entangled, substrates. Namely, we engineer composites of co-entangled microtubules and actin (46) and incorporate kinesin clusters and myosin II minifilaments that crosslink and push and pull on pairs of microtubules and actin, respectively, to generate force and motion (49, 57) (Fig 1A, SI Fig S1). To investigate the extent to which actomyosin and kinesin-microtubule activities act synergistically or antagonistically to dictate dynamics, we perform experiments with either kinesin (K, Fig 1B) or both kinesin and myosin (K+M, Fig 1B). We further characterize the effect of passive crosslinking of the microtubules (MT XL, Fig 1) or actin (Actin XL, Fig 1) at crosslinker:protein molar ratios $R$ that are high enough to induce measurable changes in the viscoelastic properties, but low enough to prevent filament bundling (47). To observe the dynamics of our active cytoskeleton composites, we collect sequential two-color time-series of actin and microtubules comprising composites over the ~1 hr time course of measurable active dynamics. As shown in Fig 1B, by simply incorporating or omitting myosin motors and passive crosslinkers, we are able to drive substantial changes in the active restructuring, emergent dynamics, and programmable phase space of non-equilibrium properties (SI Movies S1-S3).

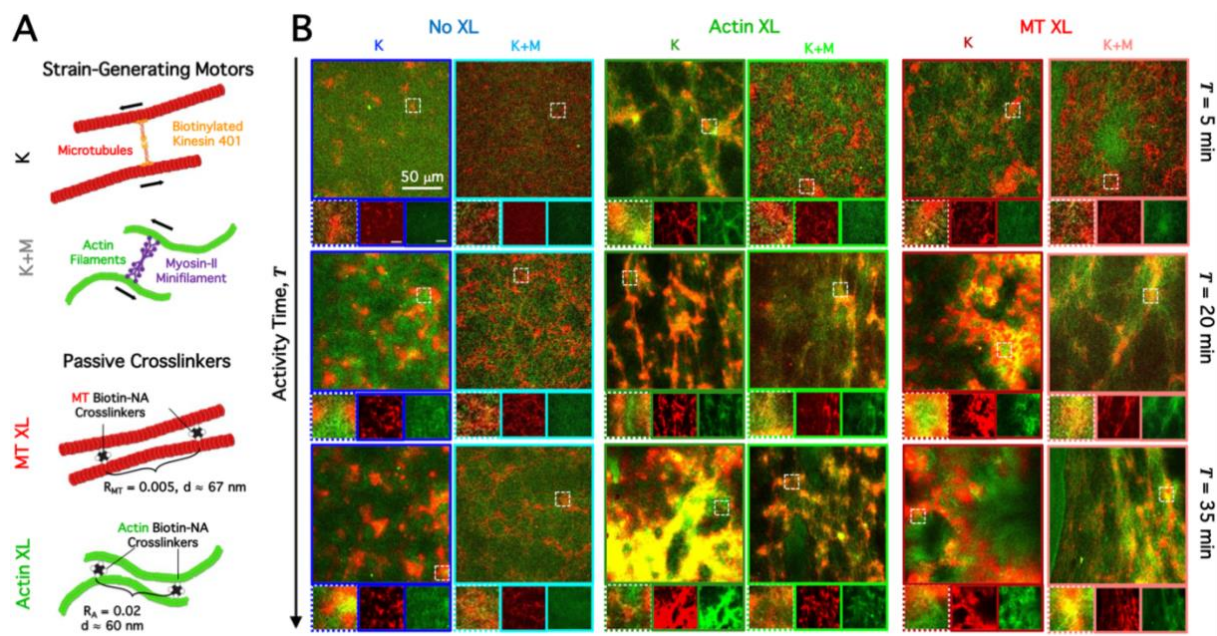

**Figure 1. Engineering and characterizing active cytoskeleton composites with varying strain-generating components and connectivity. A.** We co-polymerize actin monomers (2.32 µM) with tubulin dimers (3.48 µM) to form co-entangled composite networks of actin filaments (green) and microtubules (red). We use NeutrAvidin to passively crosslink biotinylated actin filaments (Actin XL) or microtubules (MT XL) at crosslinker:protein molar ratios of $R_A = 0.02$ and $R_{MT} = 0.005$ for actin and microtubules, to achieve similar distances $d$ between crosslinks (48). We incorporate kinesin clusters (orange) and myosin-II minifilaments (purple) to drive composites out of steady-state. **B.** We acquire two-color confocal time-series of actin (green) and microtubules (red) to capture motor-driven dynamics and reconfiguration. Each column includes images taken at three different time points, $T = 5, 20, 35$ min, during motor activity for composites with: kinesin (K, columns 1,3,5, darker shade borders), kinesin and myosin (K+M, columns 2,4,6, lighter shade borders), no crosslinking (No XL, blue hues, columns 1,2), actin crosslinking (Actin XL, green hues, columns 3,4), and microtubule crosslinking (MT XL, red hues, columns 5,6). Below each composite image is a zoom-in of a 25 µm × 25 µm region denoted by a dashed-line box in the main image, and single-channel images showing separately the microtubules (middle, red) and actin (right, green). The 50 µm scale bar in the top right panels apply to all full-size images.

All composites begin in similar structural states with interpenetrating networks of actin and microtubules uniformly distributed throughout the field of view (Fig 1B, top row). However, each composite formulation reconfigures into distinct structural states over activity times of $T \approx 30$ mins, where $T = 0$ is defined as the time at which kinesin is added to the composite. While we do not visualize the motors, the spatially uniform active dynamics that we see at the onset of activity indicates that, like the filaments, the motors are uniformly mixed throughout the composite.

Examining the three kinesin-only composites (no myosin), we find that without passive crosslinkers composites form loosely-connected MT-rich amorphous clusters. Actin filaments first co-localize in the cluster centers, but are then squeezed out into the surrounding space as the clusters contract further and disconnect from one another (Fig 1B, dark blue boxes). Passive actin crosslinking hinders this separation of actin and MTs, enabling the slow uptake of actin into MT-rich clusters, such that the composite becomes a connected network of clusters of co-localized actin and microtubules (Fig 1B, dark green boxes). Microtubule crosslinking leads to similar amorphous MT clustering and actin-MT de-mixing as without crosslinking; but these MT-rich regions coalesce over time, resulting in larger scale phase separation of actin and MTs compared to the non-crosslinked case (Fig 1B, dark red boxes).

Turning to the double-motor composites that incorporate both kinesin and myosin, we find that the addition of myosin impedes the kinesin-driven de-mixing described above and reduces the degree of restructuring over the course of activity (Fig 1B, light shaded boxes). This effect can be seen in the images at all time points (rows in Fig 1B), in which actin and microtubule networks are more evenly distributed and interpenetrating than composites without myosin. Without passive crosslinkers, composites show little rearrangement (Fig 1B, light blue boxes), as seen in previous experiments on myosin-driven actin-MT composites (26–28). Crosslinking of actin or microtubules enables more restructuring of the double-motor composites, but this reconfiguration and de-mixing is still more subdued than that for kinesin-only composites (Fig 1B, light green and red boxes).

***Actin and microtubules exhibit three distinct classes of coordinated ballistic dynamics.*** To determine the non-equilibrium dynamics that enable this rich formulation-dependent restructuring, i.e., how the system gets from one structural state to another, we perform differential dynamic microscopy (DDM) on the actin and microtubule channels of each two-channel (i.e., two-color) video. As we describe in Methods and SI, DDM analyzes differences of images separated by varying lag times $\Delta t$ in Fourier space to compute image structure functions $D(\vec{q}, \Delta t)$ for different wavevectors $\vec{q}$, which describe how density fluctuations become decorrelated for a given spatial scale (i.e., $2\pi/q$) (Fig 2A,B).

Figure 2A shows two-dimensional image structure functions $D(q_x, q_y, \Delta t)$ computed for the microtubule and actin channels of three videos that are representative of different types of dynamics we measure, which we describe below. The plots in the left and right columns correspond to $D(q_x, q_y)$ for sample 'short' and 'long' lag times, $\Delta t = 3$ s and $\Delta t = 20$ s, and the color is set by the value of $D(q_x, q_y)$, with low (blue) and high (red) values indicating lower and higher correlation, respectively (see SI Fig S2 for more $D(q_x, q_y)$ examples). The first notable feature in Fig 2A (and SI Fig S2) is the similarity in the functional form of $D(q_x, q_y)$ for actin and MT channels of the same video and lag time, indicating that the actin and MT network dynamics are well-coupled despite cases in which we observe large-scale de-mixing (Fig 1B). The lower magnitudes of actin plots compared to MTs is due to the comparatively lower signal of the actin channel. Moreover, the more uniform $D(q_x, q_y)$ values seen in the purple-bordered plots labeled '*Slow*', as compared to the middle (orange, *Fast*) and bottom (magenta, *Multimode*), are indicative of more homogeneous and slow motion, in which fluctuations decorrelate less over a given lag time and over varying lengthscales (i.e., wave vectors). The modest radial asymmetry seen most clearly in the orange-bordered plots, is a sign of anisotropic motion, which we discuss in later sections. Finally, the reduced correlation at $\Delta t = 20$ s compared to $\Delta t = 3$ s, indicates that the decorrelation timescales are <20 s.

To quantify the dynamics represented in Fig 2A, we azimuthally-average each $D(q_x, q_y, \Delta t)$ to compute a corresponding one-dimensional function for each lag time, $D(q, \Delta t)$. Figure 2B shows sample $D(q, \Delta t)$ curves for the three videos analyzed in Figure 2A. The similarity of $D(q, \Delta t)$ between actin and MT fluorescence channels indicates that actin and MT network dynamics are well-coupled, despite apparent de-mixing apparent in Fig 1B. We use the distinct functional features of each of these curves to organize our data for all composite formulations and activity times into three classes: *Slow* (top), *Fast* (middle) and *Multimode* (bottom). *Slow* curves show a monotonic, slow rise to plateau at large lag times (Fig 2B, top panel); *Fast* curves exhibit oscillations in the decorrelation plateau (Fig 2B, middle panel), and *Multimode* curves reveal two distinct plateaus at well-separated lag times (Fig 2B, bottom panel).

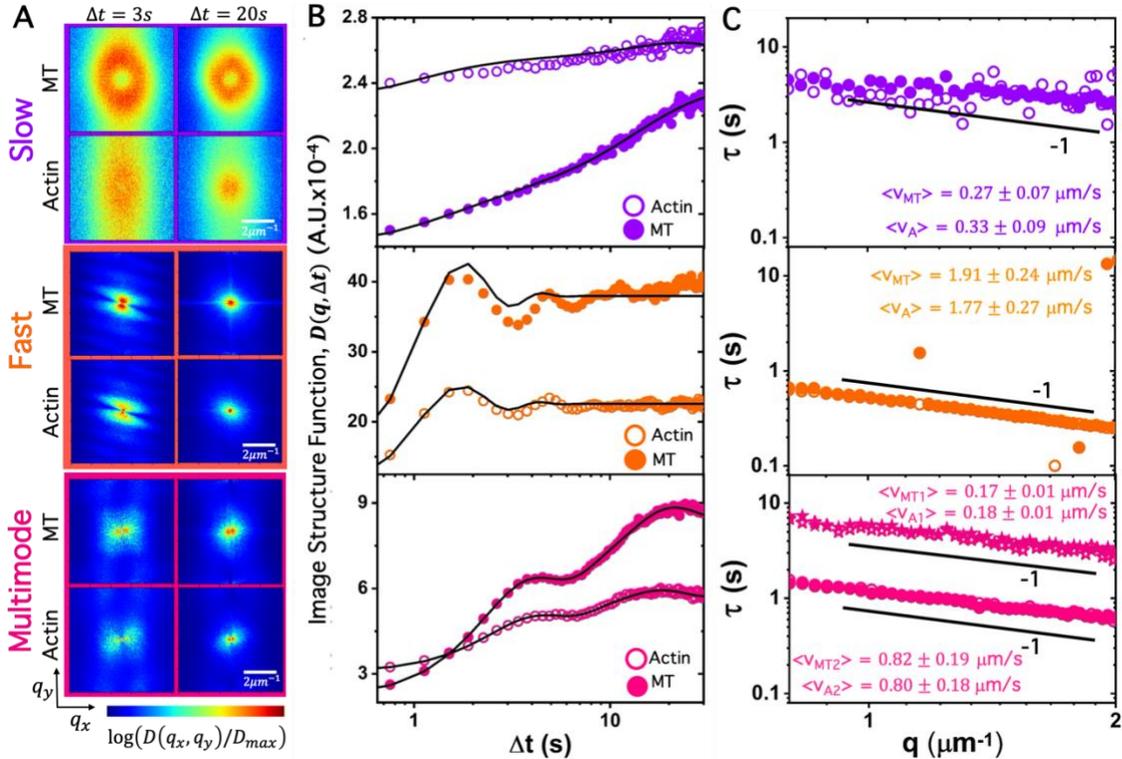

**Figure 2. Differential dynamic microscopy reveals ballistic dynamics of composites that separate into three dynamically distinct classes. A.** Representative two-dimensional image structure functions $D(q_x, q_y)$ computed from the ensemble-average of all $\Delta t = 3$ s (left) and $\Delta t = 20$ s (right) lag times for three representative videos (see SI Movies S1-S3). Color scale is normalized separately for each image, and indicates the value of each image structure function $[D(q_x, q_y, \Delta t)/D_{max}]$, with blue/red values indicating low/high correlation. **B.** Azimuthally-averaged image structure functions $D(q, \Delta t)$ versus lag time $\Delta t$ computed from 2D $D(q_x, q_y, \Delta t)$ functions for microtubules (closed symbols) and actin (open symbols) at wavevector $q = 1.33 \, \mu m^{-1}$. Black lines are fits to functions with Schulz speed distributions. **C.** Corresponding decay times $\tau(q)$ computed from $D(q, \Delta t)$ fits, universally follow $\tau(q) = (vq)^{-1}$ scaling, indicative of ballistic motion. Speeds for actin ($v_A$) and microtubules ($v_{MT}$) determined from fitting the data to $\tau(q) = (vq)^{-1}$ are listed. Listed error values are the standard deviation of the corresponding Schulz speed distribution.

These non-trivial functional forms cannot be accurately described by exponential functions typically used in DDM (26,28,58,59), so we instead use a function that assumes Schulz distributions of speeds, as has been used in other ballistic biological systems such as swimming E. coli (60,61) (see Methods and SI). This function captures the oscillatory plateaus seen in the *Fast* class, and a sum of two Schulz speed distributions accurately captures the two-plateau behavior of the *Multimode* class.

From the $D(q, \Delta t)$ fits, we extract the decay times, $\tau(q)$, which exhibit a power-law dependence on $q$ that further quantifies the type and rate of motion (Fig 2C). Despite the varied functional forms of the $D(q, \Delta t)$ data shown Fig 2B, the corresponding $\tau(q)$ curves for all classes approximately follow power-law scaling $\tau(q) \sim q^{-1}$, indicative of ballistic motion for both actin and microtubules. Similar ballistic-like dynamics have been previously reported for myosin-driven cytoskeleton composites (26, 28). Fitting each $\tau(q)$ curve to the power-law relation $\tau(q) \simeq (\langle v \rangle q)^{-1}$ yields the average speed $\langle v \rangle$ of each filament type measured over the course of the corresponding video. As listed in Fig 2C, we find that $\langle v \rangle$ for the *Fast* class is ~7× larger than the *Slow* $\langle v \rangle$. Fitting the *Multimode* $D(q, \Delta t)$ data results in two distinct $\tau(q)$ curves with corresponding $\langle v \rangle$ values that differ ≳4-fold, suggesting that *Multimode* composites undergo a combination of *Slow* and *Fast* dynamics.

In the following sections, we use the distinct $D(q, \Delta t)$ characteristics described above to correlate the *Slow, Fast* and *Multimode* classes of dynamics with composite formulation and activity time. Namely, we define the *Slow* class as having $D(q, \Delta t)$ curves that exhibit single, steady large-$\Delta t$ plateaus, *Fast* curves display single large-$\Delta t$ plateaus but with pronounced oscillations, and the *Multimode* class exhibits two distinct, steady $D(q, \Delta t)$ plateaus (Fig 2B).

***Motor competition delays the onset of acceleration and suppresses multimode dynamics.*** Having identified quantitative metrics to classify network dynamics, we now determine how the dynamics vary with composite formulation and activity time $T$. We first evaluate the average actin and microtubule speeds $\langle v \rangle$ determined from the corresponding $\tau(q)$ for each time-series (5-13 per formulation) for each of the six composite formulations. Figure 3A-C shows $T$-dependent effects of crosslinking (different panels) and motors (dark vs light shades), with speeds spanning over three orders of magnitude during motor activity. Notably, as suggested by the image structure functions shown in Figure 2B, actin and microtubule speeds are well-correlated (comparing open and closed symbols) across all composites and activity times, despite the varying degrees to which they co-localize or de-mix (Fig 1B).

We find that actin and microtubules in all composites accelerate and reach a maximum speed $v_{max}$ at activity time, $T(v_{max})$ (Fig 3D), after which $\langle v \rangle$ decreases. By classifying each data point in Fig 3A as *Slow, Fast* or *Multimode* (Fig 3E), as described above, we measure the average *Slow* speed to be $\overline{\langle v \rangle}_S \simeq$ 0.15 µm/s, which is an order of magnitude slower than the *Fast* speed of $\overline{\langle v \rangle}_F \simeq 1.8$ µm/s. The average low and high speeds for *Multimode* data are comparable to those of *Slow* and *Fast* values respectively, with $\overline{\langle v \rangle}_{M1} \simeq 0.17$ µm/s and $\overline{\langle v \rangle}_{M2} \simeq 1.7$ µm/s.

We next turn to evaluating how composite programs the different dynamical classes and their dependence on activity time $T$. The average filament speed for the uncrosslinked kinesin-driven composite (no myosin) increases ~20-fold in the first $T \approx 20$ mins, transitioning from *Slow* to *Mutimode* to *Fast* dynamics (Fig 3A), reaching $v_{max} \simeq 7$ µm/s. Following this initial acceleratory period, the composite slowly decelerates over the course of ~40 mins, but never returns to dynamics classified as *Slow*. Introducing myosin substantially delays the onset of acceleratory dynamics, increasing $T(v_{max})$ by ~3-fold, but has little impact on the magnitude of $v_{max}$ (Fig 3A,D). Moreover, *Slow* dynamics dominate over more of the activity time than for the kinesin-only composite, as seen by the higher proportion of light blue versus dark blue squares in (Fig 3A).

These results indicate that *Fast* dynamics are due primarily to kinesin-driven motion, as there is minimal change in $v_{max}$ upon addition of myosin; and that myosin activity counteracts kinesin activity to delay the onset of *Fast* dynamics, rather than cooperating synergistically to amplify active dynamics. We can understand this competition as follows.

Keeping in mind that the actin and microtubules form co-entangled interpenetrating networks of comparable mesh sizes, we can assume that every actin filament is sterically interacting with several microtubules and other actin filaments and vice versa. Kinesin acts to drive microtubules together, which, in turn, attempt to pull co-entangled actin filaments with them, competing with entanglements from other actin filaments that resist kinesin-driven straining. However, because actin filaments are more flexible and relax faster than MTs, they are able to be swept up with the kinesin-driven microtubule network and then diffuse out of MT-rich clusters to maximize their entropy.

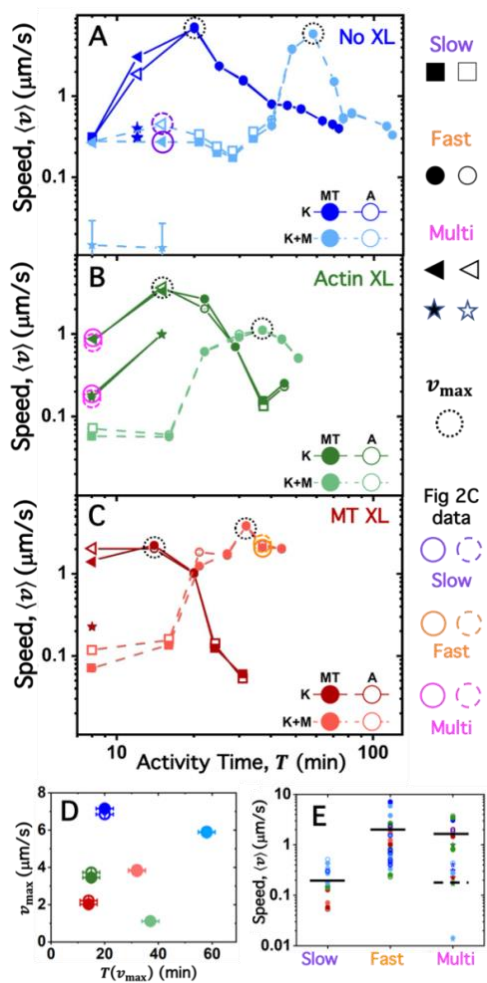

**Figure 3. Kinesin-driven composites undergo acceleration and deceleration that is gated by myosin activity and facilitated by crosslinking. A-C.** Speeds $\langle v \rangle$ of microtubules (MT, closed circles) and actin (A, open circles) versus activity time $T$ in kinesin-driven composites with no crosslinking (A, blue), actin crosslinking (B, green), and microtubule crosslinking (C, red); without myosin (K, darker shades) and with myosin (K+M, lighter shades). For *Multimode* cases, which have two speeds, the slower speed is indicated by a star. Data points corresponding to the $\tau(q)$ curves shown in Fig 2C are circled in the corresponding color (*Slow* = purple, *Fast* = orange, *Multimode* = magenta). Error bars (most too small to see) are the larger of the standard error values determined from the Schulz distribution fits and the $\tau(q)$ distributions (see Methods) **D.** Maximum speed $v_{max}$ reached by each composite, denoted by dashed circles in A, plotted against the activity time $T$ at which $v_{max}$ occurs. **E.** Scatterplot of all 106 actin and MT speeds shown in A-C, divided into *Slow, Fast* and *Multimode* classes. Horizontal lines indicate averages, with the dashed line indicating the average of the slower *Multimode* speeds (stars in A-C).

Incorporating myosin into the composites strongly enhances the competition between kinesin-driven pulling of actin and steric entanglements by pulling actin filaments together which, in turn, attempt to pull interpenetrating microtubules with them, counteracting the force of kinesin driving microtubules together. The surprising antagonistic interaction between the two motors may be due to the contractile versus extensile nature of actomyosin and kinesin-microtubule activity, respectively (62). Namely, kinesin motors are highly processive such that they principally induce nematic bundling, sliding, and extensile motion of rigid microtubules, whereas low-duty-ratio myosin motors primarily bend, compress and contract semiflexible actin filaments into asters or foci (49,57).

We expect this competition to manifest structurally as enhanced actin-microtubule mixing and interpenetration, as we see in Fig 1B. In other words, while both filament types are pulled towards like filaments (actin to actin, MTs to MTs) by their respective motors, entanglements with the other filament type resist this motor-driven self-association, thereby facilitating mixing. The net result is reduced clustering and increased actin-MT interpenetration in double-motor composites. While the dynamics eventually mirror those of kinesin-only composites, the structure remains more homogeneous, as shown in Fig 1B.

The fact that motor antagonism leads to a time-delay rather than suppression of active dynamics, suggests that eventually kinesin straining beats out myosin straining, such that the dynamics mirror kinesin-only composites, but are gated by myosin activity. Kinesin-microtubule straining likely 'beats out' actomyosin activity due to the higher density of kinesin clusters compared to myosin minifilaments. As we describe in the Methods, in all double-motor composites there are ~75 force-generating kinesin clusters for every myosin II minifilament, and the average spacing between kinesin clusters connecting a pair of microtubules is ~12 nm compared to ~2.6 μm (>$200x$ longer) for myosin minifilaments along actin filaments. This increased density of strain-generating linkers along microtubules, as well as their higher duty ratio and processivity, likely cause the kinesin-MT force-generation to dominate over that of the actomyosin.

We now turn to the effect of passive crosslinking on single-motor and double-motor composites. As shown in Fig 3A, the signatures of motor competition and activity gating seen for un-linked networks are preserved upon crosslinking of actin (Fig 3B) or microtubules (Fig 3C). The primary effect of crosslinking is a decrease in the maximum speed $v_{max}$ reached and the time over which the composites accelerate to this maximum, $T(v_{max})$ (Fig 3D). Further, both crosslinking types exhibit *Multimode* dynamics at the onset of activity (red and green triangles and stars), effectively eliminating the initial *Slow* dynamics seen in un-linked composites, likely due to crosslinking reducing the degrees of freedom and increasing the connectivity of the composites, thereby suppressing spatially uncorrelated microscale fluctuations. In other words, largescale restructuring (attributed to *Multimode* dynamics) and acceleration to $v_{max}$ is facilitated by crosslinking in kinesin-only composites. Conversely, crosslinking of double-motor composites eliminates the initial *Multimode* dynamics seen for their un-linked counterparts, instead switching directly from *Slow* to *Fast* dynamics with minimal structural reconfiguration. This reduced restructuring of crosslinked double-motor composites compared to kinesin-only composites can be seen in Fig 1.

To further understand the nature of the *Slow*, *Fast* and *Multimode* dynamics, and why crosslinking alters the propensity to exhibit each type, we return to our Fig 1 results which show that crosslinking leads to larger and denser filament aggregates compared to unlinked cases. The reduced degrees of freedom and enhanced connectivity that crosslinking provides may explain this enhanced mesoscale clustering, which, in turn, would suppress microscale fluctuations available to the more randomly distributed and less bundled filaments that emerge in the un-linked cases. This mechanistic picture suggests that *Fast* dynamics are dominated by coordinated motion or flow of the composite while uncorrelated microscale fluctuations describe the *Slow* dynamics. Conversely, as we describe above, we expect *Multimode* dynamics to arise from mesoscale restructuring, bundling and de-mixing events.

***Fast, Slow and Multimode classes correlate with distinct velocity fields and distributions.*** To corroborate the mechanisms that we postulate underlie the different dynamical classes in the preceding section, we evaluate the directionality and spatiotemporal coordination of the local dynamics that correspond to the sample *Fast, Slow* and *Multimode* data that we analyze in Fig 2.

We first create temporal color maps, which colorize each frame by the time it occurs during the video $t$, and overlay all colorized frames (Fig 4A, SI Fig S3). In this way, the maps depict the motion of the composites over the course of each video. Fig 4A shows the color maps for the actin channel, which are nearly indistinguishable from the MT channel of the same video (see Fig S3), in line with our DDM results that show that actin and MTs within any given composite exhibit similar dynamics (Figs 2,3). The *Slow* map (top panel) shows small-scale, randomly oriented motion while the *Fast* map shows spatially-coordinated and nearly unidirectional motion. The *Multimode* map displays features of both *Fast* and *Slow* dynamics, exhibiting directionality on small scales but largely uncorrelated motion at larger scales.

To quantify the features described above, we perform particle image velocimetry (PIV) on the actin and microtubule channels of the videos analyzed in Fig 4A. PIV vector fields in Fig 4B and SI Fig S4 show overlaid velocity fields at 4 equally-spaced times $t$ over the course of the videos analyzed in Fig 4A. Arrow lengths and directions represents the average velocity of features over 20 frames (~7.5 s) in the surrounding $8 \times 8$ square-pixel (6.7 μm × 6.7 μm) region of the field-of-view.

As shown in Figures 4B and S4, *Slow* fields exhibit motion that is slow (small arrows) and randomly oriented (no preferred arrow direction), while *Fast* fields show rapid directional motion with large arrows that all point in a similar direction. *Multimode* fields (Fig 4B, bottom row) reveal features of both *Slow* and *Fast* modes, as seen by the different arrow sizes and directions. Fig 4C, which shows the histograms of speeds computed from PIV analysis of each video, corroborates the dynamics we observe in the sample flow fields (Fig 4B) as well as in our DDM analysis (Fig 2C). Namely, the speed distribution for the *Fast* class (middle row) is shifted substantially to the right of that for the *Slow* video (top row), and the *Multimode* distribution (bottom row) shows two distinct peaks that approximately align with *Slow* and *Fast* distributions, respectively. To further quantify the speed distributions and compare to our DDM results, we fit each histogram to a Schulz distribution (Fig 2C solid lines), which we likewise used in the fitting function for the corresponding DDM image structure functions (see SI Methods). We find that the *Slow* and *Fast* distributions are well described by a single Schulz distribution, while the *Multimode* distributions require a sum of two Schulz distributions. The average speed $\langle v \rangle$ and standard deviation $\sigma$ determined from each fit (listed in the corresponding panel) show that the speeds measured in Fourier space using DDM (Fig 2B) and in real space using PIV are statistically indistinguishable (SI Table S2), with average values of $\overline{\langle v \rangle}_S \approx$ 0.3 μm/s, $\overline{\langle v \rangle}_F \approx 1.7$ μm/s, $\overline{\langle v \rangle}_{M1} \approx 0.2$ μm/s and $\overline{\langle v \rangle}_{M2} \approx 0.8$ μm/s for *Slow* (S), *Fast* (F), and *Multimode* (M1,M2) videos, respectively.

Motivated by the apparent class-dependent anisotropy (or lack thereof) in the PIV vector fields, we also evaluate the corresponding velocity orientation distributions (Fig 4D) which reveal isotropic motion for the *Slow* class, with no perceptible peak and comparable occurrences of all angles, compared to sustained unidirectional *Fast* motion, as evidenced by the sharply peaked narrow distribution. The *Multimode* distribution displays features of both *Fast* and *Slow* distributions, with a broader sampling of directions compared to *Fast* but with more pronounced peaks compared *Slow*.

As noted in the previous section, we also see evidence of anisotropic dynamics in our DDM analysis, manifested as radial asymmetry in the $D(q_x, q_y, \Delta t)$ plots for the *Fast* class and to a lesser extent in the *Multimode* plots (Fig 2A, SI Fig S2). To quantify this anisotropy in Fourier space we compute an anisotropy factor $A_F(q,t)$ by computing weighted azimuthal integrals of the DDM image structure function (detailed in Methods and SI (62)). $A_F$ can assume values between -1 and 1 for $x$- and $y$- directed motion, respectively, with $A_F = 0$ indicating isotropic motion. Fig 4E shows that the distribution of $A_F$ values for *Slow* and *Fast* classes exhibit distinct peaks at $A_F \approx 0$ and $A_F > 0$, indicative of isotropic and $y$-oriented motion, respectively. Conversely, the *Multimode* distribution is broader with multiple peaks that span from $A_F < 0$ to $A_F > 0$ and include a significant fraction of near-zero values. Likewise, the *Multimode* PIV orientation distribution samples a broad range of angles (isotropic, $A_F = 0$) while also exhibiting distinct peaks (directionality, $|A_F| > 0$ ).

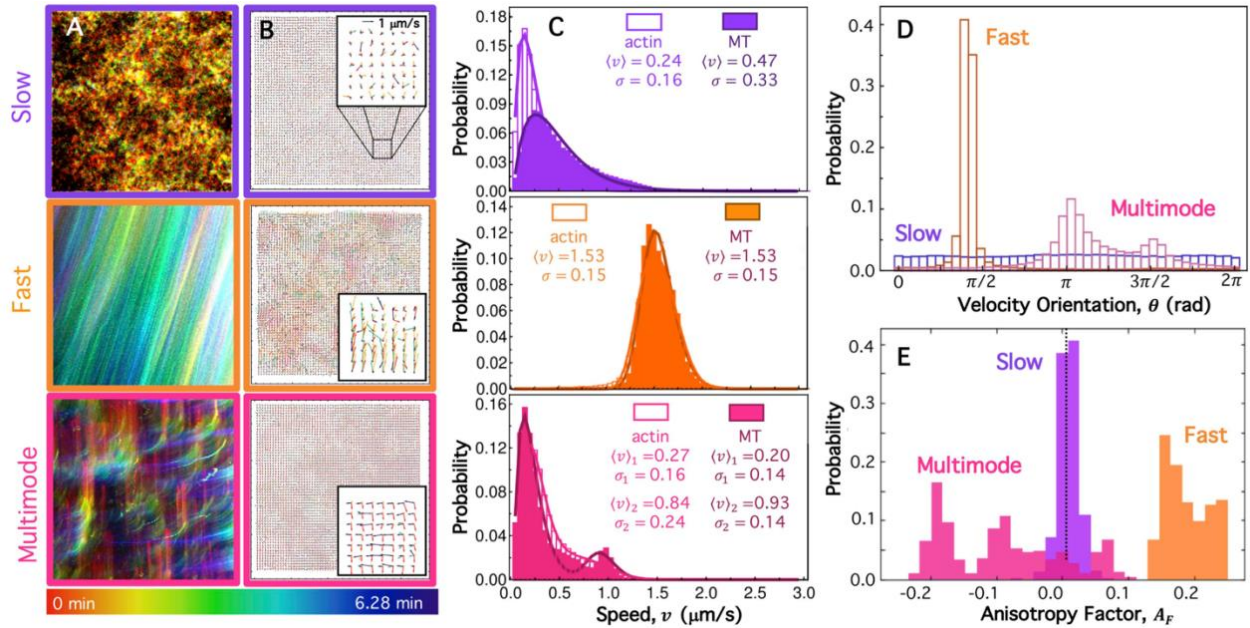

**Figure 4.** *Slow*, *Fast* and *Multimode* dynamics classified via DDM exhibit distinct PIV velocity fields and distributions. **A.** Temporal color maps which colorize the features in each 213 µm × 213 µm frame according to the time $t$ the frame is captured during the video, as indicated by the colorscale ($t = 0$ min (red) to $t = 6.28$ min (purple)), depict the motion of the composites. *Slow* (top, purple border), *Fast* (middle, orange border), and *Multimode* (bottom, magenta border) color maps correspond to the actin channel of the representative videos analyzed in Fig 2. **B.** Particle Image Velocimetry (PIV) performed on the videos analyzed in A, determines corresponding velocity vector fields in which each arrow represents the average velocity $\vec{v}(t)$ for an 8×8 square-pixel region of the 213 µm × 213 µm field. PIV vector fields for $t = 0$ s (red), 125 s (yellow), 251 s (green) and 377 s (purple) are overlaid in each panel. Insets are zoom-ins of 25 µm × 25 µm regions as indicated in the top panel. **C.** Probability distributions of speeds, determined via PIV, across all vectors in all frames of the actin (open bars) and MT (filled bars) channels of the videos analyzed in B. The solid lines are fits of each distribution to a Schulz distribution function, which describes the functional form of speed distributions assumed in DDM analysis. The average speed $\langle v \rangle$ and standard deviation $\sigma$ determined from each fit is listed in the corresponding panel in units of µm/s. The *Multimode* distribution (bottom panel, magenta) is fit to a sum of two Schulz distributions with different $\langle v \rangle$ and $\sigma$ values listed in the panel. **D.** Probability distributions of the velocity orientations that correspond to the *Slow* (purple), *Fast* (orange) and *Multimode* (magenta) actin speed distributions shown in C. **E.** Probability distributions of anisotropy factors $A_F$ computed from instantaneous DDM image structure functions $D_i$ for the same data analyzed in A-D. Dashed vertical line at $A_F = 0$ indicates isotropic dynamics whereas $A_F > 0$ and $A_F < 0$ correspond to motion in the $y$-direction (~$\pi/2$) and $x$-direction (~$\pi$), respectively. All probability distributions are dimensionless and reflect the probability that a value of the speed (C), velocity orientation (D), or anisotropy factor (E) is within the range specified by the width of the corresponding bar.

***Spatiotemporal variations in dynamics are suppressed by motor antagonism***. To better elucidate the mechanisms dictating the different dynamical classes, and the influence of motor antagonism on said mechanisms, we use both DDM and PIV to resolve variations in the short-time dynamics of the composites, i.e., those that occur within the time $t$ of a given video.

We first evaluate the average speed $\bar{v}(t)$ as a function of time $t$ for the actin and microtubule channels of each video analyzed in Figs 2 and 4, which we compute from the corresponding PIV vector fields, where $\bar{v}$ is averaged over all vectors in a single field. As shown in the $\bar{v}(t)$ plots in Fig 5A, *Slow* and *Fast* dynamics are largely stationary over the course of a given video, with nearly constant speeds. In contrast, the *Multimode* traces show discrete and abrupt shifts from intermediate to fast motion to steady slow motion.

Observing the time dependence of the corresponding average velocity orientations $\bar{\theta}(t)$, we find similar trends as for $\bar{v}(t)$, whereby the directionality of the both *Fast* and *Slow* examples is nearly independent of $t$, while the average orientation of *Multimode* vectors undergoes an abrupt and discrete shift at $t \simeq 60$ s.

To corroborate and better characterize the apparent stationarity of *Slow* and *Fast* class dynamics and the non-stationary *Multimode* dynamics shown in Fig 5A,B, we compute instantaneous DDM image structure functions $D_i(q, \Delta t, t)$, which, unlike the $D(q, \Delta t)$ curves shown in Fig 2B, are not averaged over time $t$ (62). For a system such as Brownian particles or one where the dynamics are smooth and continuous, one would expect the values of $D_i$ at a particular $q$-value and $\Delta t$ and across all times $t$ to be symmetric about $D(q, \Delta t) = \langle D_i(q, \Delta t, t) \rangle_t$. However, samples exhibiting intermittent or temporally varying dynamics may exhibit a skewed distribution of $D_i$ across times $t$. Therefore, we evaluate the probability distribution of $D_i(q, \Delta t)$ values for all $t$ in a given video to determine the extent to which dynamics are temporally heterogeneous during the acquisition time. Ergodic stationary dynamics are expected to yield Gaussian distributions of our correlation function, $D_i(q, \Delta t, t)$. As shown in Fig 5C, the *Fast* and *Slow* distributions are strongly overlapping, with the *Slow* distribution being well fit to a Gaussian function. Conversely, the *Multimode* distribution is distinctly non-Gaussian–with no obvious peak, a broad distribution of values, and significant noise–indicative of large intermittent fluctuations in structural correlation (62, 63).

To quantify the extent to which the temporal $D_i$ distributions deviate from Gaussianity, we compute the skewness $S_K = (\langle D_i - D \rangle)^3 / (\langle (D_i - D)^2 \rangle)^{3/2}$, which is zero for a Gaussian distribution. For reference, the distributions shown in Fig 5C have skewness values of $S_{K,S} \simeq 0.42$, $S_{K,F} \simeq 0.66$, and $S_{K,M} \simeq 0.86$ for the *Slow*, *Fast* and *Multimode* classes, respectively. Positive skewness, largest for *Multimode* distributions, has been reported for colloidal gels that are en route towards arrest, and has been interpreted as arising from discrete restructuring processes such as coalescing or rupturing, as well as intermittent fluctuations and rearrangements (62).

To determine the prevalence of non-stationary dynamics across the formulation phase space and activity times, we compute skewness values for all composite formulations and times $T$ evaluated in Figure 3. Fig 5D-I shows stacked confidence ellipse plots comparing skewness $S_K$, average speeds $\langle v \rangle$ and anisotropy factors $A_F$ colorized by dynamical class and separated into panels for kinesin-driven composites without (Fig 5D,F,H) and with (Fig 5E,G,I) myosin. The individual points correspond to all data points shown in Fig 3 and the ellipses enclose one standard deviation around the mean. As shown, the *Multimode* data exhibit the largest skewness values, as seen by the magenta ellipses being furthest to the right in Fig 5F,H. *Fast* and *Slow* $S_K$ values are similar to one another and deviate less from zero. The higher skewness for *Multimode* data is coupled with relatively fast speeds (Fig 5F) but low anisotropy (Fig 5H). These couplings further support our interpretation that *Multimode* dynamics arise from large intermittent restructuring events which we expect to have no preferred directionality but give rise to periods of time – e.g., during a restructuring event – that exhibit fast dynamics.

Comparison of the composites driven by kinesin only (Fig 5D,F,H; darker shades) versus two motors (Fig 5E,G,I; lighter shades) reveals that the presence of myosin nearly eliminates *Multimode* dynamics, as evidenced by the lack of magenta ellipses in Fig 5E,G,I. Further, the distributions of data points for the double-motor composites generally exhibit smaller skewness values as compared to kinesin-only composites, as seen by the ellipses shifted to the left in Fig 5G,I compared to Fig 5F,H. Despite these differences, we also observe that the distributions of speeds for composites with and without myosin are not significantly distinct, as we discussed in the previous section (also see Fig 3).

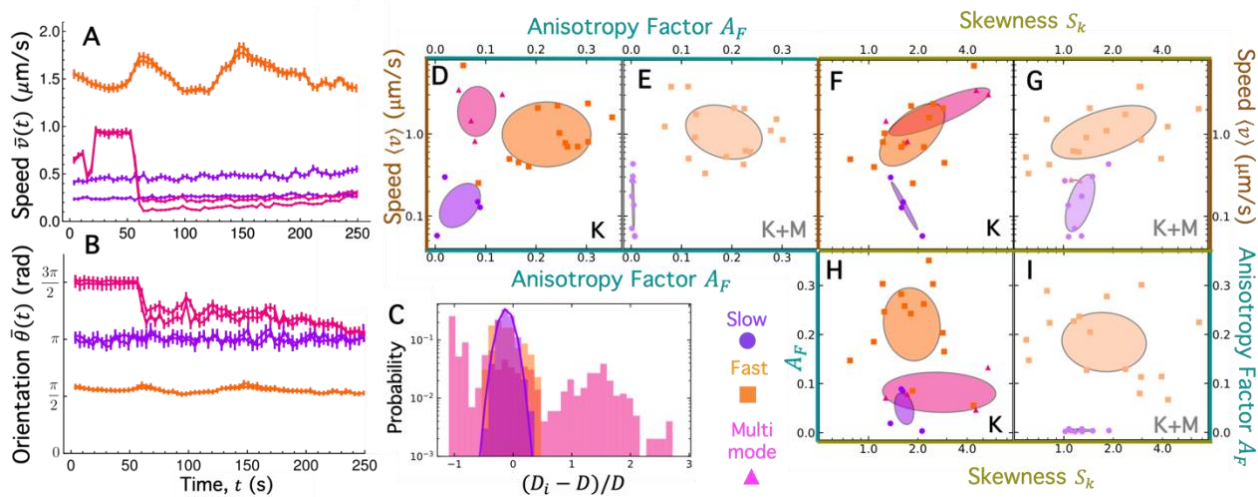

**Figure 5. Non-stationary short-timescale dynamics, unique to the *Multimode* class, indicate discrete intermittent restructuring. A**. Average speed $\bar{v}(t)$ versus time $t$ measured via PIV for the actin and MT channels of the representative *Slow* (purple), *Fast* (orange) and *Multimode* (magenta) videos analyzed in Fig 4. $\bar{v}(t)$ for each time $t$ is an average over all vector magnitudes in the PIV flow field associated with time $t$. **B**. Average velocity orientations $\bar{\theta}(t)$ versus $t$ computed from the same vector fields following the same method as in A. **C.** Probability distributions of instantaneous image structure function values, $D_i(q, \Delta t, t)$, over $t$ for $q = 0.30 \ \mu m^{-1}$ and $\Delta t \leq 3.8 \ s$ (lag times comparable to or less than typical characteristic decorrelation times) computed for the videos evaluated in A and B. To better compare the probability distributions for the different classes, $D_i$ is normalized by the $t$-averaged image structure function, $D = \langle D_i \rangle_t$. Probabilities are dimensionless and reflect the probability that a value of $(D_i - D)/D$ is within the range specified by the width of the bars. The solid purple line is a fit of the *Slow* distribution to a Gaussian function. Deviation from Gaussianity indicates sporadic discrete structural changes and is quantified by skewness $S_K$. The distributions shown for $q = 0.30 \ \mu m^{-1}$ are similar to distributions for the other measured $q$-values. **D-I**. Stacked 3-dimensional confidence ellipse plots show the relationships between average speed $\langle v \rangle$ (D-G, brown axes), anisotropy factor magnitude $|A_F|$ (D,E,H,I, teal axes), and skewness $S_K$ (F-I, gold axes). Data points, with colors and symbols indicating dynamical class according to the legend, correspond to the 106 data points plotted in Fig 3, and the ellipses enclose one standard deviation around the mean. Panels with darker shaded (D,F,H) and lighter shaded (E,G,I) ellipses display data for composites with kinesin (K) and both kinesin and myosin (K+M), respectively.

Taken together, these results demonstrate the *Multimode* dynamics arise from discrete and abrupt restructuring events and coarsening, and the presence of myosin suppresses this restructuring, such that double-motor composites exhibit very few instances of *Multimode* dynamics and remain more homogenously mixed at the end of activity. In the absence of mesoscale discrete restructuring, the double-motor networks take longer to coarsen and switch to *Fast* coordinated flow.

***Motor competition inhibits composite restructuring and de-mixing that is enhanced by crosslinking.*** To connect the dynamics we measure with various structures and reconfiguration, we develop a minimal model that aims to capture the key dynamical features of our composites. As described in Methods and SI, our model simulates filament motion that arises from motor-driven advection and thermal diffusion and works against steric hindrances and viscous traps due to motor and protein cross-linking. We purposefully simplify the model, ignoring details such as filament flexibility and individual motor dynamics that other models incorporate (64–66), to facilitate applications to other systems and identify the key parameters that dictate the experimental phenomena we observe.

Our model simulations show that all composites start as homogeneous interpenetrating networks of actin and microtubules at $T = 0$ (SI Fig S9), as we see in experiments (Fig 1B), but subsequently restructure to varying degrees depending on the composite formulation. Figure 6A, which shows sample simulation snapshots of the final states ($T = T_F$) of the six composite formulations, reveals strong suppression of restructuring by motor competition, similar to our experimental observations, with the K+M composites undergoing substantially less restructuring and de-mixing than the kinesin-only composites. Also in line with experiments, crosslinking of either actin or MTs in simulated composites enhances aggregation and clustering compared to composites without crosslinkers. This agreement between the model predictions and experimental observations suggests that it is the balance between frictional jamming and motor-driven de-mixing that dictates the different formulation-dependent structural regimes.

To quantify the degree of restructuring in simulations, we compute the probability distributions of like filaments ($g_{A/MT-A/MT}(r)$) and unlike filaments ($g_{A/MT-MT/A}(r)$) a radial distance $r$ from a given actin/microtubule (A/MT) for the initial ($T = 0$) and final ($T = T_F$) states of all composites (see SI Methods). For homogeneous, well-mixed networks, all distributions should equate to 1 for all $r$ values, which we find to be the case for the initial states of all simulated composites (SI Fig S9). The more $g_{A-A}(r)$ or $g_{MT-MT}(r)$ values are above 1 the more clustering of actin or microtubules, respectively. Conversely, $g_{A-MT}(r) < 1$ or $g_{MT-A}(r) < 1$ indicates segregation and de-mixing of actin from microtubules or vice versa. Fig 6B-E, plots the differences between the final and initial distributions, e.g. $\Delta g_{A-A} = g_{A-A}(r, T_F) - g_{A-A}(r, 0)$, such that values of zero indicate minimal restructuring while positive 'like' distributions and negative 'unlike' distributions indicate like-filament clustering and de-mixing of unlike filaments, respectively. As shown, the composites with kinesin and myosin show minimal de-mixing regardless of crosslinking ($\Delta g \approx 0$ in Fig 6C,E), while all composites without myosin show signatures of clustering and de-mixing which is generally more pronounced in the crosslinked composites (Fig 6B,D).

Finally, to directly compare the predicted and experimental restructuring, we perform identical spatial image autocorrelation (SIA) analysis (see Methods) on the initial and final experimental videos and simulation snapshots. SIA computes the correlation in intensities $g_I(r)$ between two pixels separated by a radial distance $r$ in a given image, such that $g_I(r)$ indicates the lengthscales over which structural features in an image are correlated. Specifically, $g_I(r)$ values range from 1 for complete correlation (such as when $r = 0$) and 0 for complete decorrelation, e.g., for $r$ values much larger than the size of structural features. Similar to Fig 6B-E, we evaluate the differences between final and initial correlation functions $\Delta g_I(r)$ for actin and microtubules in all simulated composites (Fig 6H,I), which we compare to experimental values (Fig 6F,G). We find that, in both experiments and simulations, the presence of myosin reduces the distance over which structural correlations are enhanced over the time-course of motor activity, evidenced as faster decay in $\Delta g_I(r)$ with increasing $r$ in Fig 6G,I compared to Fig 6F,H. This feature is indicative of reduced large-scale clustering and de-mixing of actin and microtubules, as is also evident in Figs 1B and 6A. Moreover, in both experiment and simulations, passive crosslinking generally leads to increased structural correlations (larger $\Delta g_I$ values) compared to composites without crosslinkers, in particular at larger distances and for actin crosslinking. The increased aggregation with actin crosslinking manifests in experiments as minimal decay and non-monotonic dependence of $\Delta g_I(r)$ with increasing $r$, for actin and microtubules, respectively, indicative of fewer small-scale clusters and increased mesoscale (>10 µm) structural correlations. In simulations, increased aggregation can be seen as larger $\Delta g_I$ values in the presence of actin crosslinkers across all lengthscales.

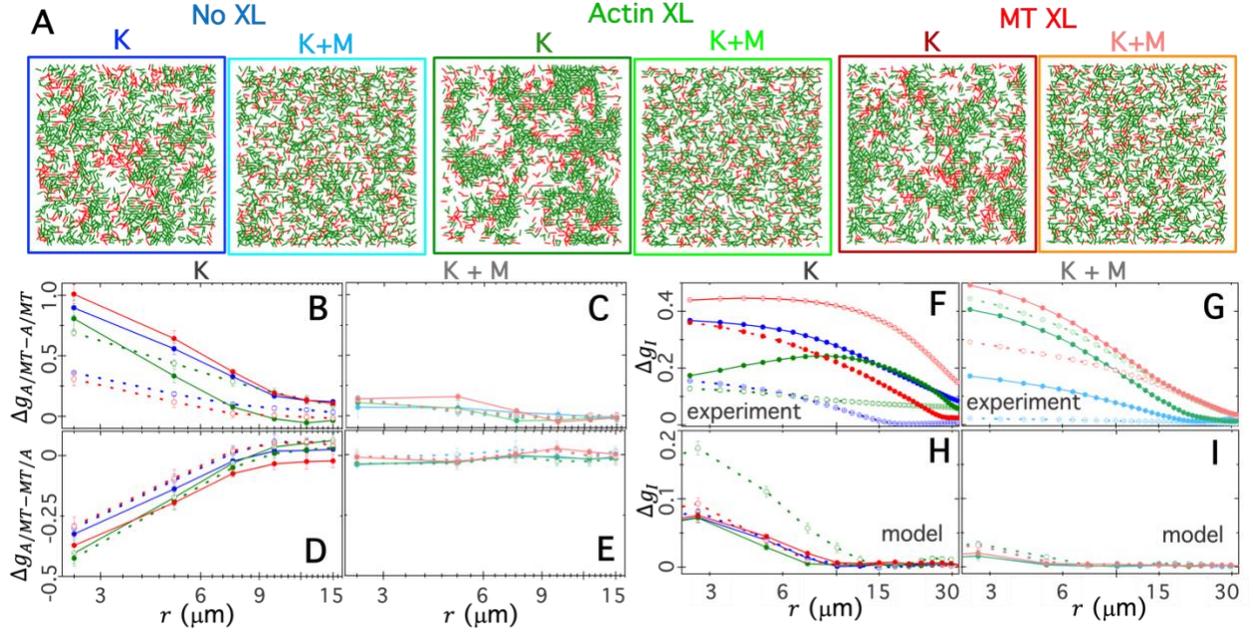

**Figure 6. Minimal advection-diffusion modeling characterizes and corroborates expected motor-driven restructuring. A.** Simulation snapshots, each 155 μm × 155 μm, show sample final configurations of microtubules (red) and actin (green) in composites with no passive crosslinking (No XL, blue borders), actin crosslinking (Actin XL, green borders) and microtubule crosslinking (MT XL, red borders), subject to active forcing by kinesin (K, darker borders) or both kinesin and myosin (K+M, lighter borders). Simulation details are provided in SI Methods and Table S1. **B-I**. The difference between the initial ($T = 0$) and final ($T = T_F$) values of various structural correlation functions, $\Delta g_\_(r) = g_\_(r, T_F) - g_\_(r, 0)$, versus radial distance $r$ between two filaments (B-E) or pixels (F-I) for composites depicted in A. The type of passive crosslinking is color-coded according to the border colors in A. **B,C**. The difference in like-filament distributions for actin ($\Delta g_{A-A}$, open symbols) and microtubules ($g_{MT-MT}$, filled symbols) in composites with (B) kinesin or (C) both kinesin and myosin. **D,E**. Unlike-filament distribution differences for (D) actin ($g_{A-MT}$, open symbols) and (E) microtubules ($g_{MT-A}$, filled symbols) for the composites analyzed in B and C. **F-I**. Spatial image autocorrelation differences $\Delta g_I$ for actin (open symbols) and microtubules (filled symbols) computed from (F,G) experimental time-series and (H,I) simulation snapshots for composites with (F,H) kinesin or (G,I) both kinesin and myosin. Error bars for simulations are standard error across 3 replicates, and for experiments are standard error across 100 images from 3 independent time-series.

We note that given the simplicity of our model and the simulated renderings of the composites, as well as the noise in our microscope images, as can be seen in Fig 1 and SI Movies S1-S3, we do not expect quantitative agreement between experiment and simulations. Rather, we aim to capture qualitatively similar dependences of structural features on crosslinking and motor competition, as we describe above. Namely, the presence of myosin inhibits restructuring while the passive crosslinking enhances it. The generally larger $\Delta g_I$ values measured in experiments compared to simulations is likely due to the noise and finite depth of the experimental images which limit the occurrence of 'empty space' seen in simulated composites, thereby overestimating correlations across lengthscales. Moreover, the flexibility of the actin filaments, not accounted for in the model, may also allow for greater restructuring and larger clusters to form.

To understand the underlying mechanisms driving this restructuring more fully, we consider that while kinesin motor activity adds to the advective term for microtubules in the model, the processive nature of kinesin also increases the drag on the microtubules. Conversely, addition of non-processive myosin motors increases filament advection with a relatively smaller increase in drag. Thus, kinesin activity acts to collect microtubules into locally arrested clusters that can either sweep up or squeeze out actin filaments. The addition of passive crosslinking of actin or microtubules accelerates this process by facilitating coalescence of smaller clusters into larger ones. On the other hand, myosin activity allows for filament redistribution within clusters and diffusive migration of filaments out of clustered regions, thereby inhibiting segregation between actin and microtubules and increasing the rate at which newly formed clusters can dissolve back into a mixed state. Succinctly stated, motor antagonism can arise from an interplay between competitive motor-driven advection and frictional drag, irrespective of its origin–steric interactions or passive or active crosslinking.

**Conclusion**

The cytoskeleton is a non-equilibrium multifunctional composite comprising diverse protein filaments, motors, and crosslinkers that cooperate and compete to enable diverse cellular structures and processes. As such, the cytoskeleton is one of the primary inspirations to the burgeoning field of active matter, and much of current active matter research seeks to learn from and emulate the cytoskeleton. The composite nature of the cytoskeleton, which confers its signature versatility and programmability, is one of its hallmarks. Yet, current active matter platforms are largely limited to a single force-generating component and/or substrate. We address this gap by engineering co-entangled and crosslinked composites of microtubules and actin filaments driven by kinesin and myosin motors–breaking new ground in active matter design by incorporating multiple independently tunable force-generating components and active substrates.

By coupling Fourier-space and real-space analyses (DDM and PIV) we show that composites undergo a combination of *Fast* advective flow, *Slow* isotropic fluctuations, and *Multimode* restructuring that result in structures ranging from interpenetrating actin-microtubule scaffolds to de-mixed amorphous clusters. Surprisingly, competition between kinesin and myosin straining delays the onset of kinesin-driven acceleration without appreciably changing the range of speeds the different composites exhibit. Motor antagonism also suppresses mesoscale restructuring events that underlie *Multimode* dynamics thereby sustaining mixed networks of actin and microtubules. Conversely, passive crosslinking hastens the onset of kinesin-mediated acceleration and subsequent deceleration by enhancing network connectivity and suppressing uncorrelated microscale motion. Importantly, the emergent dynamics and extensive programmable phase space of non-equilibrium properties we reveal are a result of very subtle changes in substrate connectivity and activity.

Our work brings reconstituted cytoskeleton systems an important step closer to mimicking the complexity of the active composite cytoskeleton by integrating two distinct and ubiquitous motor-filament systems– actomyosin and kinesin-microtubule networks–that have been shown to interact and co-mediate important cellular processes including morphogenesis and exocytosis (71,72), mechanosensation (69), and migration and stiffening (70). Interactions between actomyosin, kinesin and microtubules have also been implicated in wound healing, mitosis, and cytoplasmic streaming (7, 15, 16, 28, 43, 71–73). As the motor and filament concentrations in our composites are within physiological ranges (74), our results may offer insight into the macromolecular dynamics and interactions that contribute to these cellular processes.

For example, Ref 73 investigates the dynamics of vesicles, moving along microtubules via kinesin, and the actin mesh that surrounds the vesicles, in oocytes. The authors use PIV and DDM to demonstrate that both vesicles and actin exhibit a combination of ballistic advection and active diffusion. These multimodal dynamics, with speeds and diffusion coefficients similar between vesicles and actin, are reminiscent of the *Multimode* dynamics we measure, which are likewise similar between actin and microtubules. Ref 73 further reports that the ballistic component requires microtubules while active diffusion is dictated by the actin mesh. Similar in spirit, our results suggest that the *Fast* mode is driven by kinesin, while actomyosin activity delays this advection. Finally, Ref 73 suggests that crosslinking of actin and microtubules may play an important role in determining vesicle dynamics, speeding up the ballistic contribution, similar to the effect that crosslinking of either actin or microtubules has on the dynamics we measure.

However, there are several key differences between our results and those of Ref 73 that point to distinct mechanisms at play. We find that both modes are ballistic, rather than a diffusive and ballistic term; and the speeds we measure are at least an order of magnitude faster. This effect is likely due to the increased motor activity and reduced crowding in our system compared to Ref 73. The non-equilibrium actin dynamics in Ref 73 arise from active filament assembly and disassembly via actin binding proteins, rather than the direct push and pull of myosin motors. This difference is likely what underlies the slow ballistic motion we measure compared to active diffusion. Moreover, while Ref 73 reports that actin slows vesicle diffusion via steric constraints, we find that actomyosin activity simply delays the onset of fast kinesin-driven flow, with little impact on the measured speeds, suggestive of competition between kinesin and myosin. In addition, we measure the dynamics of microtubules being acted on by many kinesin motors, rather than vesicles that are carried by a single kinesin motor. The collective action of many kinesin motors pushing and pulling on connected microtubules likely gives rise to the faster dynamics we measure. Finally, we note that cell-like confinement of in vitro cytoskeletal networks has been shown to play a key role in recapitulating dynamics and structures seen in cells (75–77). We plan to build in this layer of complexity in our future work (78).

Beyond the biological relevance, the programmability of our composites, with multiple well-controlled tuning knobs–motors, filaments and crosslinkers–which can each be varied independently while maintaining composite integrity, opens the door for reconfigurable materials that can be programmed to exhibit varying types and rates of motion and restructuring over broad spatiotemporal scales. For example, materials based on our designs could be used as spatially-controlled micro-actuators, responsive filtration and sequestration devices, and self-curing and self-repairing infrastructure technologies. Our minimal advection-diffusion model that recapitulates our experimental trends, is broadly applicable to active composite networks, and lays the foundation for more complex predictive models that quantitatively capture the structure and dynamics of composite active matter. As such, we anticipate our double-motor material design, intriguing dynamical results, and corresponding modeling framework, will spark a new class of studies that explore the broad parameter space of this platform.

**Methods**

See SI Methods Section for more detailed descriptions of each of the following sections.

*Protein Preparation:* Rabbit skeletal actin monomers (Cytoskeleton), biotin-actin monomers (Cytoskeleton), porcine brain tubulin dimers (Cytoskeleton), biotin-tubulin dimers (Cytoskeleton), rhodamine-labeled tubulin dimers (Cytoskeleton), and myosin-II (Cytoskeleton), are reconstituted and flash-frozen into single-use aliquots according to previously described protocols (28, 48). Biotinylated kinesin-401 is expressed in Rosetta (DE3)pLysS competent *E. coli* (ThermoFisher) and purified as described in the SI.

For composites that incorporate actin or microtubule crosslinking, actin-actin or microtubule-microtubule crosslinker complexes are prepared according to previously described protocols (48). In brief, biotin-actin or biotin-tubulin is combined with NeutrAvidin and free biotin at a ratio of 2:2:1 protein:free biotin:NeutrAvidin.

Immediately prior to experiments: myosin-II is purified as previously described (27) and stored at 4°C, and kinesin clusters are formed by incubating dimers at a 2:1 ratio with NeutrAvidin with 4 µM DTT for 30 minutes at 4°C.

*Active Cytoskeleton Composite Preparation:* Actin-microtubule composites are formed by polymerizing 2.32 µM unlabeled actin monomers and 3.48 µM tubulin dimers (5% rhodamine-labeled) in PEM-100 (100 mM PIPES, 2 mM MgCl$_2$, 2 mM EGTA) supplemented with 0.1% Tween, 10 mM ATP, 4 mM GTP, 5 μM Taxol, and 0.47 μM AlexaFluor488-phalloidin (Life Technologies) to label the actin.

For crosslinked composites, a portion of either actin monomers or tubulin dimers is replaced with equivalent crosslinker complexes to achieve the same overall actin and tubulin concentrations and crosslinker:protein ratios of $R_A = 0.02$ for actin or $R_{MT} = 0.005$ for microtubules. $R_A$ and $R_{MT}$ values are chosen to achieve similar lengths between crosslinkers $d$ along actin filaments and microtubules ($d_A \simeq 60$ nm, $d_{MT} \simeq 67$ nm) (48) and to be high enough to induce measurable changes in the viscoelasticity compared to unlinked networks, but low enough to prevent filament bundling.

Composites are polymerized for 30 mins at 37°C, after which 1.86 μM phalloidin is added and the composite is incubated for 10 mins at room temperature. 50 μM blebbistatin is added to inhibit myosin-actin interaction prior to de-activation via 488 nm illumination (26), and an oxygen scavenging system is added. Finally, 0.47 μM myosin-II and 0.35 μM kinesin are added. Concentrations of actin, tubulin, myosin-II and kinesin are within reported physiological ranges (74, 79, 80).

While myosin activity is controlled by blebbistatin de-activation, kinesin starts to act immediately, so $T = 0$ for each experiment is set as the time kinesin is added. Each sample is gently flowed into a ~1 mm × 24 mm sample chamber composed of a silanized (81) coverslip and microscope slide fused together by a ~100 μm thick parafilm spacer and sealed with epoxy, creating an airtight chamber.

*Fluorescence Microscopy:* Imaging of AlexaFluor488-labeled actin and rhodamine-labeled microtubules comprising composites is performed using a Nikon A1R laser scanning confocal microscope with a 60× 1.4 NA oil-immersion objective (Nikon), 488 nm laser with 488/525 nm excitation/emission filters, and 561 nm laser with 565/591 nm excitation/emission filters. 488 nm illumination also locally de-activates blebbistatin (26–28). Time-series (videos) of 256 × 256 square-pixel (213 μm × 213 μm) images are collected at 2.65 fps for a maximum video time of $t_{max} = 1000$ frames (~377 s ≃ 6.28 mins). Imaging begins 5 mins after the addition of kinesin motors ($T = 5$ min) in the middle of the ~100 μm thick sample chamber. Each successive video is collected in a different field of view of the same sample until there is no longer any discernible restructuring or motion ($T \simeq 60 - 120$ mins). 7-15 videos are collected for each of the six formulations (no crosslinking, actin crosslinking and microtubule crosslinking; with and without myosin). Each video includes two channels that separate the actin and microtubule signals such that they can be processed separately and compared.

*Differential Dynamic Microscopy (DDM):* DDM is performed on the actin and microtubule channels of each video as described previously (28). Image structure functions are determined by taking the square of 2D Fourier transforms of differences between an image at time $t$ and one at $t + \Delta t$. This yields the instantaneous image structure function, $D_i(q_x, q_y, \Delta t, t)$ where $q_x$ and $q_y$ are $x$ and $y$ components of the wave vector $\vec{q}$. As typically done in DDM analysis, we average $D_i$ over all times $t$ (frames) of a given video, and all wave vectors $\vec{q}$ with the same magnitude $q$, to determine the 1D image structure function $D(q, \Delta t)$ that can be fit to various models. We fit $D(q, \Delta t)$ versus $\Delta t$ for each wave vector $q$ to a model in which the distribution of speeds are described by one or two Schulz functions (60) (see SI Methods), as has been used to describe other ballistic biological systems (60, 61). In cases in which one distribution is used to describe the data (*Slow* and *Fast* data), there are 4 free parameters ($A, B, \tau_1, Z_1$), whereas for *Multimode* data, there are 7 (adding $\tau_2, Z_2, f$) (see SI Methods). For each video, we perform fits for at least 40 different $q$ values in the range $q = 0.8 - 2$ μm$^{-1}$ (~3 – 8 μm), from which we extract $\tau(q)$ curves for the actin and microtubule channels. By fitting each $\tau(q)$ curve to $\tau(q) = (\langle v \rangle q)^{-1}$ we compute the average speed $\langle v \rangle$ for each channel of each video. We determine the error associated with $\langle v \rangle$ using two methods. First, we compute $v$ from each individual ($\tau, q$) pair (i.e., $v = 1/\tau q$) and determine the standard error across those values. Secondly, we use the Schulz parameter $Z$ determined from our $D(q, \Delta t)$ fits and our measured $\langle v \rangle$ value to compute the standard deviation $\sigma$ and corresponding standard error via the relation $Z = \left(\frac{\langle v \rangle}{\sigma}\right)^2 - 1$. Error bars shown in Fig 3 are the larger of the two values for each case.

To determine the degree to which dynamics deviate from radial symmetry, implying directionality, we compute the anisotropy factor $A_F$ of $D_i(q_x, q_y, \Delta t, t)$ in $q$-space by computing $A_F(q, \Delta t, t) = \int_0^{2\pi} D(q, \Delta t, \theta) \cos(2\theta) \, d\theta / \int_0^{2\pi} D(q, \Delta t, \theta) d\theta$ and averaging over $q$, $\Delta t$ and $t$ (82, 83). $\theta$ is defined relative to the positive $y$-axis such that $A_F > 0$ and $A_F < 0$ correspond to motion along the $y$- and $x$-direction, respectively, and $A_F = 0$ indicates isotropic motion.

To evaluate the time-dependence of dynamics over short timescales (within the time $t$ of a single video), we also investigate the temporal distribution of instantaneous image structure functions $D_i(q_x, q_y, \Delta t, t)$ for a given $\Delta t$ and $q$. For steady-state dynamics, one would expect this distribution to be Gaussian. Deviations from Gaussianity indicate sporadic events which cause larger than typical structural decorrelations. We quantify this non-Gaussian behavior by evaluating the skewness, $S_K = \langle (D_i - D) \rangle^3 / (\langle (D_i - D)^2 \rangle)^{3/2}$, where the average is over $\Delta t$ and $q$.

*Particle Image Velocimetry (PIV):* PIV analysis is performed using the GPU-accelerated version of OpenPIV (84). Interrogation windows of 8×8 square-pixels, with a 4×4 square-pixel overlap, are used to generate 64 × 64 grids of velocities for microtubule and actin channels of each time-series. Average velocities $\vec{v}$ for each interrogation window are determined from image pairs separated by $\Delta t = 10$ frames (~3.77 s). From the measured velocities, we determine the distribution of individual speeds $v(t)$, and velocity orientations $\theta(t)$ across each image over the course of a video. To identify and exclude spurious velocities during statistical analysis, we rejected those points for which the signal-to-noise ratio was less than 2. We fit the speed distributions to Schulz distributions by minimizing the mean square difference between the predicted statistical weight assigned to each bin (of width 50 nm/s) for a given choice of parameters and the actual fraction of speeds in each bin. Arrows plotted in Fig 4B and Fig S4 represent the local velocity on a regular Cartesian grid, with arrow length proportional to speed. Visualizations at different video times $t$ are superposed, with arrow color representing $t$.

*Spatial Image Autocorrelation (SIA):* SIA analysis is performed separately on actin and microtubule channels of microscope images and simulation snapshots (see below) using custom Python scripts (27,28,83). SIA measures the correlation in intensity $g_I$ in an image as a function of separation $\vec{r}$ (85). That is, $g_I(\vec{r}) = \langle I(\vec{r'} + \vec{r}) I(\vec{r'}) \rangle_{\vec{r'}}$ where $\vec{r'}$ represents the position of each pixel within an image. We perform an azimuthal average to generate $g_I(r)$ where $r$ is the magnitude of $\vec{r}$. Computationally, this autocorrelation function is found by taking the Fourier transform of the image, multiplying by its complex conjugate, and applying an inverse Fourier transform. To normalize the autocorrelation functions so that the maximum is 1 (i.e., $g_I(0) = 1$), we subtract from each image the mean of that image and divide the image by its standard deviation before performing the Fourier transforms. Correlation curves shown in (i) Fig 6F,G and (ii) Fig 6H,I, are averages across (i) 100 microscope images from 3 independent time-series and (ii) simulation snapshots from 3 independent runs (see below). Error bars indicate standard error.

*Computational Model:* To predict motor-driven composite restructuring, we develop a minimal model that captures the key energetic components of our system, as fully described in the SI Methods and Table S1. In brief, we allow filaments to interact with neighboring filaments via (i) motor-generated forces that can either pull the interacting filaments towards each other or push them away; or (ii) crosslinks that increase the friction forces on the interacting filaments, as described in SI Ref 18. The movement of a filament center to a neighboring grid point within a small temporal time step is then a stochastic event whose probability can be calculated by the standard solution to the Fokker-Planck equation given by: $p_i(x \geq l) = 1 - \frac{1}{2}(1 + \text{erf}(\frac{l - \mu_i}{\sigma_i \sqrt{2}}))$, where $l$ is the distance to the next grid point in a particular direction, $\mu_i$ is the average advection induced displacement in that direction and $\sigma_i$ is the diffusion-based rms 1D displacement of the filament along the direction to the specific grid point. The subscript $i$ represents a specific filament in the model. The movement probability of filament $i$ to a neighboring grid point that (i) contains filament $j$'s center or (ii) is empty is given by (i) $p_{ij} = p_i \times p_j$ or (ii) $p_{ij} = p_i \times 1$.

We implement our model on a 150 µm x 150 µm hexagonal lattice with 2.5 µm spacing, and use numerical values for all model parameters that are based on experimental and literature values (see SI Table S1). Initially, each lattice point is either empty or occupied by a microtubule center or actin filament center using probabilities matching the average volume fraction occupied by these elements. The movement of the filaments is simulated in each iteration by calculating the likelihood of each possible movement $p_{ij}$ for all grid points $i$ containing filament centers, and randomly picking one of these movements to occur based on these probabilities (Fig S8) (87). We perform three independent simulation runs for each composite formulation (SI Fig S9).

To quantify the degree of clustering and segregation of the different filaments, we compute the probability distributions of filaments that are alike, $g_{A-A}(r) = <\frac{N_A(r)}{f_A N(r)}>$ or $g_{MT-MT}(r) = <\frac{N_{MT}(r)}{f_{MT} N(r)}>$, and unlike, $g_{A-MT}(r) = <\frac{N_{MT}(r)}{f_{MT} N(r)}>$ or $g_{MT-A}(r) = <\frac{N_A(r)}{f_A N(r)}>$ a radial distance $r$ from a given actin filament (A) or microtubule (MT) (see SI Methods). In the above, $N_{A/MT}(r)$ is the number of actin/microtubule neighbors at distance $r$ from a specific filament, $f_{A/MT}$ is the actin/microtubule volume fraction, and $N(r)$ is the maximum number of neighbors possible a distance $r$ from the specific actin/microtubule. An increase in $g_{A/MT-A/MT}(r)$ above 1 indicates clustering of actin/microtubules, while a decrease in $g_{A/MT-MT/A}(r)$ below 1 indicates segregation of unlike filaments. Correlation analysis data shown in Fig 6B-E are averages over all filaments of the same type over three statistically independent replicates with error bars representing the standard error.

**Data Availability:** All data generated or analyzed during this study are included in this published article and its supplementary information files, or available from the corresponding author on reasonable request.

**Competing Interest Statement:** The authors declare no competing interests.

**Author Contributions:** RMRA conceived the project, guided the experiments, interpreted the data, and wrote the manuscript. RJM analyzed and interpreted the data and wrote the manuscript. KAL purified and assayed the kinesin. JLR helped conceive the project, guide experiments, interpret data, and write the manuscript. PK and CG designed and performed modeling and simulations. MD helped interpret data, develop theoretical models, and write the manuscript. CC performed experiments and analyzed data. JM and MS analyzed data and prepared figures. MJR helped guide experiments and provide useful feedback.

**Acknowledgments:** We thank Leila Farhadi, Maya Hendija, Nadia Schwartz Bolef and Lauren Melcher with assistance in developing kinesin purification protocols, performing image analysis and developing simulations. This research was funded by a WM Keck Foundation Research Grant and NSF DMREF Award (DMR 2119663) awarded to RMRA., JLR., MD, and MJR; and an NIH R15 Award (NIGMS award R15GM123420 awarded to RMRA and RJM). JYS acknowledges startup support from the Keck Science Department of Scripps, Pitzer, and Claremont McKenna Colleges.

**Motor antagonism dictates emergent dynamics in active double networks tuned by crosslinkers**

*Ryan J. McGorty[1], Christopher J. Currie[1], Jonathan Michel[2], Mehrzad Sasanpour[1], Christopher Gunter[3], K. Alice Lindsay[4], Michael J. Rust[5], Parag Katira[3], Moumita Das[2], Jennifer L. Ross[4], Rae M. Robertson-Anderson[1,\*]*

[1]Department of Physics and Biophysics, University of San Diego, San Diego, California 92110, USA
[2]School of Physics and Astronomy, Rochester Institute of Technology, Rochester, New York 14623, USA
[3]Department of Mechanical Engineering, San Diego State University, San Diego, CA 92182, USA
[4]Department of Physics, Syracuse University, Syracuse, New York 13244, USA
[5]Department of Molecular Genetics and Cell Biology, University of Chicago, Chicago, Illinois 60637, USA
*randerson@sandiego.edu*

**Supplementary Information**

**Contents:**

**SI Methods:** Detailed descriptions of all methods and materials.

**Table S1:** Parameters used in mathematical model and simulations.

**Table S2:** Comparison of average speeds $\langle v \rangle$ and corresponding standard deviations $\sigma$ measured with PIV and DDM.

**Movie S1:** Four sample time-series of actin-microtubule composites exhibiting *Slow* dynamics.

**Movie S2:** Four sample time-series of actin-microtubule composites exhibiting *Fast* dynamics.

**Movie S3:** Four sample time-series of actin-microtubule composites exhibiting *Multimode* dynamics.

**Figure S1:** Cartoon of phase space of different composite formulations.

**Figure S2:** Two-dimensional and azimuthally-averaged DDM image structure functions for 9 additional time-series with *Slow*, *Fast*, and *Multimode* dynamics.

**Figure S3:** Temporal color maps for 9 time-series with *Slow*, *Fast*, and *Multimode* dynamics.

**Figure S4:** PIV vector fields for 9 additional time-series with *Slow*, *Fast*, and *Multimode* dynamics.

**Figure S5:** Fits of 9 additional *Slow*, *Fast*, and *Multimode* speed distributions to Schulz functions.

**Figure S6:** Time course of average filament speeds and orientations for 9 additional time-series with *Slow*, *Fast*, and *Multimode* dynamics

**Figure S7:** Stacked 3-dimensional confidence ellipse plots show the relationships between average speed $\langle v \rangle$, anisotropy factor $A_F$, and skewness $S_K$ for different composite formulations.

**Figure S8:** Sample plot showing simulation mechanics.

**Figure S9:** Simulation snapshots for three independent trials for each composite formulation.

**Supplementary Methods:**

*Protein Preparation:* Rabbit skeletal actin monomers (Cytoskeleton, AKL99, Lot#139), biotin-actin monomers (Cytoskeleton, AB07, Lot#49), porcine brain tubulin dimers (Cytoskeleton, T240, Lot#121), biotin-tubulin dimers (Cytoskeleton, T333P, Lot#27), rhodamine-labeled tubulin dimers (Cytoskeleton, TL590M, Lot#31), and myosin-II (Cytoskeleton, MY02, Lot#19), are reconstituted and flash-frozen into single-use aliquots according to previously described protocols (1).

Biotinylated kinesin-401 is expressed in Rosetta (DE3)pLysS competent *E. coli* (ThermoFisher) grown on selective media plates for 16-18 hours at 37°C. Fifteen colonies are added to a 5 ml starter culture of selective LB media and grown for 2 hours at 37°C/250rpm before adding to 400 ml of selective LB media. Cells are grown at 37°C/250rpm to OD 0.6-0.9 at 600 nm, then induced at 20°C/250rpm for 18 hours with 1mM Isopropyl β-D-1-thiogalactopyranoside (IPTG), and pelleted at 5,000 rpm for 10 minutes at 4°C before being frozen at -80°C for 1 hour. Cells are lysed in lysis binding buffer (50 mM PIPES, 4 mM MgCl$_2$, 20 mM imidazole, 10 mM β-mercaptoethanol, 50 μM ATP, one protease inhibitor tablet per 10 ml, 1.1mg/ml PMSF, 1.1mg/ml lysozyme) via sonication for 3 mins, pulsing every 20 seconds, then pelleted for 30 mins at 40,000 x g at 4°C, filtered through a 0.22 uM filter, and incubated with 1 ml nickel (Ni-NTA) agarose beads (Qiagen) for 2 hours on a rocker at 4°C. The lysate/bead mixture is passed through a chromatography column then washed with 15 ml buffer (50mM PIPES, 4mM MgCl2, 20mM imidazole, 10 mM β-mercaptoethanol, 50 μM ATP, one protease inhibitor tablet per 10ml) before 1 mL fractions are eluted in (50 mM PIPES, 4 mM MgCl2, 20 mM imidazole, 10mM β-mercaptoethanol, 50 μM ATP, one protease inhibitor tablet per 10 ml, 2 mM DTT, 0.05 mM ATP). An elution dot blot is performed to assess the most concentrated fraction which is run through a 40K MWCO desalting column for buffer exchange with PEM-100 with 0.1mM ATP, then mixed with 60% sucrose for a final concentration of 10% sucrose before being aliquoted and flash-frozen into single-use aliquots.

For composites that incorporate actin or microtubule crosslinking, actin-actin or microtubule-microtubule crosslinker complexes are prepared according to previously described protocols (2). In brief, biotin-actin or biotin-tubulin is combined with NeutrAvidin and free biotin at a ratio of 2:2:1 protein:free biotin:NeutrAvidin.

Immediately prior to experiments: myosin-II is purified as previously described (1) and stored at 4°C, and kinesin clusters are formed by incubating the dimers at a 2:1 ratio with NeutrAvidin (ThermoFisher) with 4 μM DTT for 30 minutes at 4°C.

*Active Cytoskeleton Composite Preparation:* Actin-microtubule composites are formed by polymerizing 2.32 μM unlabeled actin monomers and 3.48 μM tubulin dimers (5% rhodamine-labeled) in PEM-100 (100 mM PIPES, 2 mM MgCl$_2$, 2 mM EGTA) supplemented with 0.1% Tween, 10 mM ATP, 4 mM GTP, 5 μM Taxol, and 0.47 μM AlexaFluor488-phalloidin (Life Technologies, A12379) to label the actin.

For crosslinked composites, a portion of either the actin monomers or the tubulin dimers is replaced with equivalent crosslinker complexes to achieve the same overall actin and tubulin concentrations and crosslinker:protein ratios of $R_A = 0.02$ for actin or $R_{MT} = 0.005$ for microtubules. $R_A$ and $R_{MT}$ values are chosen to achieve similar lengths between crosslinkers $d$ along actin filaments and microtubules ($d_A \simeq 60$ nm and $d_{MT} \simeq 67$ nm). As previously described (2), we estimate these values using $d_A = \frac{l_{monomer}}{2R}$, where $l_{monomer}$ is the length of an actin monomer, and $d_{MT} = \frac{l_{ring}}{26R}$, where $l_{ring}$ is the length of a ring of 13 tubulin dimers. Crosslinking ratios are also tuned to be high enough to induce measurable changes in the viscoelastic properties compared to unlinked networks, but low enough to prevent filament bundling (2).

Actin and tubulin concentrations are chosen to be similar to those used in previous studies on myosin-driven actin-microtubule composites (1, 3, 4), and such that the mesh sizes for the actin and microtubule networks are comparable ($\zeta_A \simeq 0.96$ μm and $\zeta_{MT} \simeq 1.44$ μm, respectively), and the effective composite mesh size is $\zeta_C \simeq (\zeta_A^3 + \zeta_{MT}^3)^{-1/3} \simeq 0.64$ μm (5). Further fine-tuning of the concentrations is achieved through a series of optimization experiments to identify a formulation space in which composites reliably form percolated networks and are visibly active on the timescale of minutes.

Composites are polymerized for 30 mins at 37°C, after which 1.86 μM unlabeled phalloidin is added and the composite is incubated for 10 mins at room temperature. 50 μM blebbistatin is added to inhibit myosin-actin interaction prior to de-activation via 488 nm illumination (1), and an oxygen scavenging system (45 μg/mL glucose, 0.005% β-mercaptoethanol, 43 μg/mL glucose oxidase, 7 μg/mL catalase) is added to reduce photobleaching. Finally, 0.47 μM myosin-II and 0.35 μM kinesin (pre-formed into complexes) are added. Concentrations of actin, tubulin, myosin-II and kinesin in composites are within reported physiological ranges of ~2.6 – 70 μM, ~1.3 – 19 μM, 0.4 – 4.8 μM, and 0.1 – 1.6 μM, respectively (6).

While myosin activity is controlled by blebbistatin de-activation, kinesin starts to act on microtubules immediately, so the start of the activity time of each experiment, $T_A = 0$, is set as the time kinesin is added. Each sample is gently flowed into a ~1 mm ($x$) × 24 mm ($y$) sample chamber composed of a silanized (7) coverslip and microscope slide fused together by a ~100 μm thick parafilm spacer and sealed with epoxy, creating an airtight chamber. We see no visible signs of sample drift from leaking or heating as our control systems (no motors) display no discernible bulk motion or restructuring. We do note that in cases in which motors induce directional motion, the motion is preferentially along the long '$y$' axis of the chamber. We do not expect this preferred directionality to artificially bias any other structural or dynamical features of the composite.

*Fluorescence Microscopy:* Imaging of AlexaFluor488-labeled actin and rhodamine-labeled microtubules comprising composites is performed using a Nikon A1R laser scanning confocal microscope with a 60× 1.4 NA oil-immersion objective (Nikon), a 488 nm laser with 488/525 nm excitation/emission filters, and a 561 nm laser with 565/591 nm excitation/emission filters. 488 nm illumination also locally activates myosin-II ATPase activity by de-activating blebbistatin as previously described (1, 3, 4). Time-series (videos) of 256 × 256 square-pixel (213 μm × 213 μm) images are collected at 2.65 fps for 1000 frames ($t = 0 - 377$ s $\simeq 6.28$ mins). Acquisition of the first video for each sample starts 5 mins after the addition of kinesin motors ($T = 5$ min) in the middle of the ~100 μm thick sample chamber. Each successive video is collected in a different field of view of the same sample until there is no longer any discernible restructuring or motion ($T \simeq 45 - 120$ mins). 7-15 videos, each spanning acquisition times of $t = 0 - 377$ s, are collected for each of the six composite formulations (no crosslinking, actin crosslinking and microtubule crosslinking; with kinesin and with kinesin and myosin).

*Differential Dynamic Microscopy (DDM):* DDM is performed separately on the actin and microtubule channels of each 1000-frame video using custom written python scripts as described previously (1, 4). Image structure functions are determined by taking the square of 2D Fourier transforms of differences between an image at time $t$ and one at $t + \Delta t$. This yields the instantaneous image structure function, $D_i(q_x, q_y, \Delta t, t_v)$ where $q_x$ and $q_y$ are $x$ and $y$ components of the wave vector $\vec{q}$. As typically done in DDM analysis, we average $D_i$ over all times $t$ (frames) of a given video, and all wave vectors $\vec{q}$ with the same magnitude $q$, to determine the 1D image structure function $D(q, \Delta t)$ that can be fit to various models. We fit $D(q, \Delta t)$ versus $\Delta t$ for each wave vector $q$ to the sum of either one or two Schulz speed distributions:

$$A\left(1 - \left(\left[f\left(\frac{\tau_1(Z_1+1)}{Z_1 * \Delta t} * \frac{\sin(Z_1 * \arctan(\theta_1))}{(1+\theta_1^2)^{\frac{Z_1}{2}}}\right)\right] + \left[(1-f)\left(\frac{\tau_2(Z_2+1)}{Z_2 * \Delta t} * \frac{\sin(Z_2 * \arctan(\theta_2))}{(1+\theta_2^2)^{\frac{Z_2}{2}}}\right)\right]\right)\right) + B,$$

where amplitude $A$, background $B$, decay times $\tau_1$ and $\tau_2$, amplitude fraction $f$, and Schulz numbers $Z_1$ and $Z_2$ are $q$-dependent free parameters, and $\theta_n = \frac{\Delta t}{\tau_n(Z_n+1)}$ (8). Schulz numbers characterize the speed distributions $P(v) = \frac{v^Z}{Z!}\left(\frac{Z+1}{\bar{v}}\right)^{Z+1} \exp\left[-\frac{v(Z+1)}{\bar{v}}\right]$ where $Z = \left(\frac{\bar{v}}{\sigma}\right)^2 - 1$. We use the functional form of $D(q, \Delta t)$ and the corresponding Schulz distribution fits to divide our data into three dynamical classes:

*Slow*, *Fast* and *Multimode*. *Slow* data as those with $D(q, \Delta t)$ curves that exhibit a single flat decorrelation plateau and are well-fit to a single Schulz distribution (i.e., $f = 1$). *Fast* data are classified by $D(q, \Delta t)$ curves that are also well-fit to a single Schulz distribution, but have decorrelation plateaus that exhibit pronounced $\Delta t$-dependent oscillations. *Multimode* data have $D(q, \Delta t)$ curves that display two distinct plateaus and are best fit to the sum of two Schulz distributions with comparable $f$ values. $\tau(q)$ curves for each composite and time $T$ are extracted from the corresponding $D(q, \Delta t)$ fits.

For *Slow* and *Fast* data, in which one distribution describes the data, there are 4 free parameters ($A, B, \tau_1, Z_1$). For *Multimode* data, this number increases to 7 (adding $\tau_2, Z_2, f$). For each video these fits are performed over 40 different $q$ values in the range $q = 0.8 - 2$ µm$^{-1}$ (~3 – 8 µm), from which we extract $\tau(q)$ curves for the actin and microtubule channels of each of the 7-15 statistically different videos we collect for each of the six composite formulations.

Unreliable fits to the data would result in noisy $\tau(q)$ curves or curves that do not display power-law behavior over the entire $q$ range. On the contrary, we find that all composites for all times $T$ during activity exhibit $\tau(q) \sim q^{-1}$ scaling indicative of ballistic motion from which we compute the average speed $\langle v \rangle$ by fitting to $\tau(q) = (\langle v \rangle q)^{-1}$. We determine the error associated with the measured $\langle v \rangle$ using two methods. First, we compute $\langle v \rangle$ from each individual $(\tau, q)$ pair (i.e., $v = 1/\tau q$) and determine the standard error of the distribution of those values. Secondly, we use the Schulz parameter $Z$ determined from our $D(q, \Delta t)$ fits and our measured $\langle v \rangle$ to compute the standard deviation $\sigma$ and corresponding standard error via the relation $Z = \left(\frac{\bar{v}}{\sigma}\right)^2 - 1$. The error bar for each data point in Fig 3 represents the larger of the two standard error values.

All composites exhibit $\tau(q) \sim q^{-1}$ scaling indicative of ballistic motion (9) and the average speed $\langle v \rangle$ is computed by fitting $\tau(q)$ to $\tau(q) = (\langle v \rangle q)^{-1}$. Error bars shown in Fig 3 represent the standard error of the distribution of speeds computed from each individual $q$ value (i.e., $v = 1/\tau q$) in the range over which we fit $D(q, \Delta t)$.

To determine the degree to which dynamics deviate from radial symmetry, implying directionality, we compute the anisotropy factor $A_F$ of $D_i(q_x, q_y, \Delta t, t)$ in $q$-space by computing $A_F(q, \Delta t, t) = \int_0^{2\pi} D(q, \Delta t, \theta) \cos(2\theta) \, d\theta / \int_0^{2\pi} D(q, \Delta t, \theta) d\theta$ and averaging over $q$, $\Delta t$ and $t$ (10, 11). $\theta$ is defined relative to the positive $y$-axis such that $A_F > 0$ and $A_F < 0$ correspond to motion along the $y$- and $x$-direction, respectively, and $A_F = 0$ indicates isotropic motion.

To evaluate the time-dependence of dynamics over short timescales (within the time $t$ of a single video), we also investigate the temporal distribution of instantaneous image structure functions $D_i(q_x, q_y, \Delta t, t)$ for a given $q$. For steady-state dynamics, one would expect this distribution to be Gaussian. Deviations from Gaussianity indicate sporadic events which cause larger than typical structural decorrelations. We quantify this non-Gaussian behavior by evaluating the skewness, $S_K = \langle (D_i - D) \rangle^3 / (\langle (D_i - D)^2 \rangle)^{3/2}$, where the average is over $\Delta t$ and $q$.

*Particle Image Velocimetry (PIV):* PIV analysis is performed using the GPU-accelerated version of Open-PIV (12). We use interrogation windows of $8 \times 8$ square-pixels, with a $4 \times 4$ square-pixel overlap, to generate $64 \times 64$ grids of velocities for both the microtubule and actin channel of each time-series. Average velocities $\vec{v}$ for each interrogation window are determined from image pairs separated by $\Delta t = 10$ frames (~3.77 s), with the starting frame for each successive interval overlapping with the ending frame for the previous interval. From the measured velocities, we determine the distribution of individual speeds $v(t)$, and velocity orientations $\theta(t)$ over the course of a video. Because of the heterogeneous spatial distribution of fluorescent material, the signal-to-noise ratio of estimated velocities varied appreciably. To identify and exclude spurious velocities during statistical analysis, we rejected those points for which the signal-to-noise ratio was less than 2. To fit Schulz distributions for flow speed, we first partitioned velocities into bins of width 50 nm/s, computing the fraction of velocities in each bin. Schulz distribution parameters were then chosen by minimizing the mean square difference between the predicted statistical weight assigned to each bin for a given choice of parameters and the actual fraction of speeds in each bin. To visualize velocity fields using vector plots, we smoothed vector fields to eliminate spurious vectors in two steps. First, we removed vectors with unsatisfactory signal-to-noise ratios, and replaced the velocity vectors at the corresponding locations by local mean method, as implemented in the OpenPIV Spatial Analysis Toolbox. In the local mean approach, an invalid vector is iteratively replaced by the mean of all valid velocity vectors in a local patch centered about the site of the spurious vector. Here, we use a 3×3 averaging region. If at some location, no valid vectors are available at adjacent grid sites for a given iteration, a velocity field is not computed. The process is repeated until all spurious vectors are replaced. Next, we removed velocities that had a component that was more than 2 standard deviations greater than the global mean, and replaced them by the local mean method described above. Arrows plotted in Fig 4B and Fig S3 represent the local velocity on a regular Cartesian grid, with arrow length proportional to speed. Visualizations at different video times $t$ are superposed, with arrow color representing $t$.

*Spatial Image Autocorrelation (SIA)*: SIA analysis is performed on the actin and microtubule channels separately for each frame of each video using custom Python scripts, previously validated for similar active systems (4, 13, 14). SIA measures the correlation in intensity $g_I$ of two pixels in an image (video frame) as a function of separation distance $r$ (13). We generate autocorrelation curves $g_I(r)$ by taking the fast Fourier transform of the image $F(I)$, multiplying by its complex conjugate, applying an inverse Fourier transform $F^{-1}$, and normalizing by the squared intensity: $g_I(r) = \frac{F^{-1}(|F(I(r))|^2)}{[I(r)]^2}$. We radially average $F(I)$ to compute a single average correlation value for each lengthscale $r$ of a given image, independent of direction. We use a spatial resolution of 1 pixel (832 nm) and perform SIA over the entire $256 \times 256$ square-pixel $(213\ \mu m)^2$ image. We also perform SIA on skeletonized versions of the same images to reduce potential noise from introducing artifacts. The trends we observe in skeletonized and raw images are statistically indistinguishable. Finally, we perform this same SIA analysis on images from simulations (see Figs 6, S9). Correlation curves shown in (i) Fig 6F,G and (ii) Fig 6H,I, are averages across (i) 100 microscope images from 3 independent time-series and (ii) simulation snapshots from 3 independent runs (see below). Error bars indicate standard error.

*Computational Model:* To predict the restructuring of the composites due to motor activity, we develop a minimal model, based on the framework described in Ref 18 and references therewithin, that captures the key energetic components of the composites. We define the available space as a hexagonal grid with periodic boundary conditions. Each grid point can be occupied by an actin or microtubule filament center or can be empty. The filaments can interact with neighboring filaments within reach, via 1) motor-generated forces that can either pull the interacting filaments towards each other or push them away from each other; or 2) crosslinks that increase the friction forces on the interacting filaments. The movement of a filament center to a neighboring grid point within a small temporal time step is then a stochastic event whose probability can be calculated by the standard solution to the Fokker-Planck equation given by

$p_{ij}(x \geq l) = 1 - \frac{1}{2}\left(1 + \mathrm{erf}\left(\frac{l - \mu_{ij}}{\sigma_i \sqrt{2}}\right)\right)$ … (1),

where $l$ is the distance to the next grid point $j$ in a particular direction, $\mu_{ij}$ is the average advection induced displacement from the current grid location in the direction from $i$ to $j$, and $\sigma_i$ is the diffusion-based root mean squared (rms) 1D displacement of the filament along the direction to the specific grid point. The subscript $i$ represents a specific filament in the model and $j$ represents a neighboring grid point. The average advection-induced displacement along a given direction, $\mu$, is a function of time elapsed since the filament moves to the new grid point, and is calculated as

$$\mu_{ij}(t) = \frac{(\sum_{j \neq i} F_{ij}) \cdot \hat{\eta}}{\gamma_i} \Delta t + \mu_{ij}(t - \Delta t) \quad \ldots (2),$$

where $F_{ij}$ is the force from the motors between filament $i$ and a same-type (actin or microtubule), interacting filament $j$, given by the force per motor ($F_m$ or $F_k$ for myosin or kinesin, respectively) times the number of motors per filament ($N_m$ or $N_k$). The direction of $F_{ij}$ is along the line joining the two filament centers and can be attractive or repulsive depending on filament orientations; $\hat{\eta}$ is the unit vector along the direction of motion to the neighboring grid point; and $\Delta t$ is the time-step for which the probability of motion is being calculated. $\gamma_i$ is the effective friction factor, given by the sum of the viscous drag on filament $i$, ($\gamma_A$ or $\gamma_{MT}$ for actin or microtubule, respectively), the protein friction from all motors between interacting filaments of similar type ($\gamma_m * N_m * N_{A,i}$ or $\gamma_k * N_k * N_{MT,i}$) and the protein friction from crosslinks between similar type filaments ($\gamma_X * N_{A,i}$ or $\gamma_X * N_{MT,i}$). $\gamma_m$ ($\gamma_k$) is the friction factor per myosin (kinesin) motor between two filaments, $N_{A,i}$ ($N_{MT,i}$) is the number of actin (microtubule) filaments interacting with the current actin (microtubule) filament, and $\gamma_X$ is the crosslinker friction factor between each interacting filament of similar type.

The diffusion based rms displacement of a filament in a specific direction is calculated using

$$\sigma_i = \sqrt{2D(t_i + \Delta t)},$$
$$D = \frac{k_B T}{\gamma_i} \quad \ldots (3),$$

Where $t_i$ is the time a filament has been in grid location $i$, $k_B$ is the Boltzmann constant, and T is the temperature of the system.

The movement of a filament center to a neighboring grid point occupied by another filament center is restricted sterically and can be only accomplished if the two filaments exchange positions. Thus, in such a scenario, the cumulative movement probability of filament $i$ to a neighboring grid point containing filament $j$'s center is given by

$$p_{ij,c} = p_{ij} \times p_{ji} = p_{ji,c} \quad \ldots (4),$$ which is the same for the filament at grid point $j$ exchanging its location with filament at $i$.

Within the same spirit, the movement of a filament from grid point $i$ to a neighboring grid point $j$ that is empty can by calculated as

$$p_{ij,c} = p_{ij} \times 1 \quad \ldots (5).$$

We purposefully choose a minimal approach to capture the dynamics to shed light on the competing factors of motor activity and friction from crosslinkers. Our model assumes a single length for all filaments while in experiments actin and microtubules assume a distribution of lengths. We treat all filaments as rigid rods while actin in experiments is semiflexible with a persistence length of ~17 μm. Our simulations are in 2D while experimental composites span 3D. Our future work will build these additional features into our model.

Some important justifications and derivations that underlie our modeling approach include:

1. While crosslinkers are often thought of as springs connecting different filaments, individual crosslinker bonds are reversible and transiently switch between bound and unbound states. When a force is applied on this system, either internally via molecular motors, or via an external force (such as tension or shear), crosslinkers can slip along the length of the filaments as they transiently bind and unbind. The rate at which the filaments slip past each other (or past crosslinks on neighboring filaments) gives an estimate of the viscosity of the system. This slip also results in plastic deformation or yielding in the materials. A simplified molecular theory of viscosity, based on breaking of elastic bonds, slippage, and bond reformation between neighboring elements of a macroscale system, such as a crosslinked polymer network, is well described by Ref 18. The model relates describes the slipping rate $v$ as a function of the force $F$ driving this slip by the equation $v = F/(Nfk_s\tau_{on})$, where $F$ is the driving force, $N$ is the number of crosslinkers, $f$ is the duty ratio (fraction of time the bonds are bound), $k_s$ is the elastic stiffness (spring constant) of the linker, and $\tau_{on}$ is the average bond lifetime. This simplified approximation provides a good estimate of viscous, irreversible, rearrangements that can occur in crosslinked polymer networks with reversible crosslinker binding. The term in the RHS denominator is a measure of the viscosity of the system or the viscous drag on individual filaments. We want to once again clarify that the elastic nature of crosslinks is not being ignored, just being coupled with the transient nature of the crosslinker bonds. If $\tau_{on}$ is really large, i.e. the crosslinker bonds are really strong, then $v$ will be really small, so there will be almost no permanent slip or plastic deformation. In this limit, the network will have an elastic response to external force.

2. Under the simplified assumptions of the Ref 18 model description, when an external force is trying to slip two crosslinked filaments past each other, this force is instantaneously balanced by the elastic force developed in the crosslinkers connecting the two filaments $F = k_s x$, where $x = vt$ is the average displacement between the two filaments, $v$ is the slip velocity, and $t$ is the average elapsed time. If an individual crosslinker bond has a finite average lifetime of $\tau_{on}$, then the maximum force that is resisted by each crosslinker bond is $F = k_s v \tau_{on}$. If $N$ is the total number of crosslinkers, and $f$ is the average time a crosslinker stays bound, then the force balance becomes $F = Nfk_s v\tau_{on}$. Under the simplified assumption that both $f$ and $\tau_{on}$ are constant, the resistive force is then proportional to the sliding velocity and the rest of the terms on the RHS can be treated as an effective drag on the filaments. While this assumption is indeed simplistic, as $f$ and $\tau_{on}$ will change based on the force applied, in the regime we are considering, where the external forces per crosslinker are smaller than the characteristic dissociation force of the crosslinker bond, this assumption holds. The application of this force-velocity relationship has been used effectively to describe filament sliding and force generation characteristics of actin-myosin and microtubule-kinesin systems, as reviewed in Ref 18 and references therewithin.

3. We estimate the drag coefficient of individual myosin on actin and kinesin on microtubules from the RHS of the force equation $F = Nfk_s v\tau_{on}$. $N$ is the number of myosins or kinesins interacting between actin and microtubule filaments, which is 1 for the drag from individual motors, but is included in the calculation of the total effective drag (see explanation for equation 2). $f$ is calculated as $k_{on}/(k_{on} + k_{off})$ where $k_{on}$ and $k_{off}$ are the binding and unbinding rates of molecular motors with their respective filaments. Using the rates given in Ref 15 for skeletal muscle, the motor protein stiffness of 4 pN/nm based on estimates from Ref 18, and using $k_{off}^{-1} = \tau_{on}$, we estimate the drag per myosin on actin as ~0.2 pN.ms/nm. The value given for the drag coefficient in Ref 15 for actin sliding due to myosin activity is 0.4 pN.ms/nm. Both values are much higher than the fluid drag on the filament, as reported in Ref 17. Also, since they only balance the active force generation by attached myosin motors and any external force on the filament with this drag force, the drag coefficient should include the passive effects of bound motors, which are a combination of molecular friction as described above and fluid drag. A value in the range of 0.2-0.4 pN.ms/nm for myosin drag on actin is in line with a 2 pN force from single myosin motors moving actin filaments at speeds of ~5-10 μm/s. Similarly, drag coefficients for kinesin motors are calculated from duty ratio and off rates given in Ref 18 giving a value ~6 pN.ms/nm, which can be compared to the 5 pN force generated per motor and ~0.8 μm/s microtubule sliding speeds.

We implement our model on a 155 μm x 155 μm 2D space with a hexagonal lattice, where the lattice spacing is 2.5 μm. Initially, each lattice point is either occupied with a microtubule filament center, an actin filament center or is left empty using probabilities matching the average volume fraction occupied by these elements. The movement of the filaments is simulated in each iteration by calculating the likelihood of each possible movement, $p_{ij,c}$ for all grid points $i$ and $j$, where at least one of them contains a filament center, and randomly picking one of these movements to occur based on these probabilities. Since each movement occurs over a timescale of $\Delta t$, the effective time progression per movement can be approximated by $\frac{2\Delta t}{\sum_{i,j} p_{ij}}$ ($i \neq j$, at least $i$ or $j$ are occupied by a filament center) at each iteration step. We take the value of $\Delta t$ as 0.1 s, at the start, such that both the rms displacement due to diffusion and the average advection distance due to motor driven forces are both smaller than the grid size, but dynamically adjust the $\Delta t$ at the next iteration step to match the effective time progression from the previous iteration step. The simulation is run for $T_S = 10^6$ iterations, which we find is sufficient to reach quasi-steady state. Specifically, running the simulation for $0.5T_S$, $0.8T_S$, $T_S$, and $1.2T_S$ iterations, we observe insignificant change in the filament distributions for $\geq 0.8T_S$. The model calculations and simulations are coded in python and the scripts are available on GitHUB (https://github.com/compactmatterlab/active-filament-networks). A cartoon depiction of the model is shown in Fig S8 and numerical values for all model parameters are included in Table S1. We perform three independent simulation runs for each composite formulation for error analysis (see Fig S10).

To quantify the degree of clustering and segregation of the different filaments, we compute the probability distributions of like ($g_{A-A}(r)$, $g_{MT-MT}(r)$) and unlike ($g_{A-MT}(r)$, $g_{A-MT}(r)$) filaments a radial distance $r$ from a given actin filament (A) or microtubule (MT) as:

$g_{A-A}(r) = <\frac{N_A(r)}{f_A N(r)}>$ and $g_{MT-MT}(r) = <\frac{N_{MT}(r)}{f_{MT} N(r)}>$ for like filaments, and

$g_{A-MT}(r) = <\frac{N_{MT}(r)}{f_{MT} N(r)}>$ and $g_{MT-A}(r) = <\frac{N_A(r)}{f_A N(r)}>$ for unlike filaments.

In the above, $N_{A/MT}(r)$ is the number of actin/microtubule neighbors at distance $r$ from a specific filament, $f_{A/MT}$ is the volume fraction of actin/microtubules in the simulation space, and $N(r)$ is the maximum number of possible neighbors a distance $r$ from the specific actin/microtubule filament. An increase in $g_{A/MT-A/MT}(r)$ above 1 indicates clustering of actin/microtubules, while a decrease in $g_{A/MT-MT/A}(r)$ below 1 indicates segregation of actin/microtubules from microtubules/actin. We perform correlation analysis up to $r = 15$ μm which we found sufficient to capture most of the correlation decay with $r$. Large radial distances display periodicity due to the periodic boundary conditions incorporated into the model. Spatial analysis algorithms also exclude filaments located at the maximum radial analysis distance or less from the simulation boundaries, to prevent the boundaries from skewing the results. Correlation analysis data shown in Fig 6 are averages over all filaments across three statistically independent replicates with error bars representing the standard error.

|  | Description | Value | Reference |
|---|---|---|---|
| **Total Grid Size** |  | 150 μm x 150 μm |  |
| **% actin filaments** | % of 2D space taken up by actin filament | 40% | experimental |
| **% microtubules (MT)** | % of 2D space taken up by MTs | 15% | experimental |
| **Grid spacing ($l$)** |  | 2.5 μm |  |
| **Filament length** | Length of each actin filaments and MT | 5 μm | experimental |
| $F_m$ | Force generated per myosin motor | 3 pN | (15) |
| $F_k$ | Force generated per kinesin motor | 6 pN | (16) |
| $N_m$ | Number of myosin motors per actin-actin interaction | 10 | experimental |
| $N_k$ | Number of kinesin motors per actin-actin interaction | 5 | experimental |
| $\Delta t$ | Time increment | 100 ms |  |
| $\gamma_A$ | Viscous drag on an actin filament | 0.005 pN·ms/nm | (17) |
| $\gamma_{MT}$ | Viscous drag on a microtubule filament | 0.01 pN·ms/nm | (18) |
| $\gamma_m$ | Viscous drag on the filament due to single myosin motor binding | 0.3 pN·ms/nm | (15) |
| $\gamma_k$ | Viscous drag on the filament due to single kinesin motor binding | 6 pN·ms/nm | (18) |
| $\gamma_X$ | Viscous drag on the filament due to single cross-linker binding | 10 pN·ms/nm | lower-bound (19) |
| T | Temperature of the system | 290 K |  |

**Table S1: Parameters used in mathematical model and simulations.** Specific numerical values of parameters are chosen to match experimental conditions, including the concentrations of actin, microtubules, motors and crosslinkers. Values for motor forces and viscous drag terms are based on literature values as specified in the table.

| Class | Channel | Method | | | |
|---|---|---|---|---|---|
|  |  | PIV | | DDM | |
|  |  | ⟨v⟩ | σ | ⟨v⟩ | σ |
| **Slow** | microtubule | 0.48 | 0.33 | 0.27 | 0.07 |
|  | actin | 0.23 | 0.17 | 0.33 | 0.09 |
| **Fast** | microtubule | 1.49 | 0.18 | 1.91 | 0.24 |
|  | actin | 1.60 | 0.21 | 1.77 | 0.23 |
| **Multimode** | microtubule - v1 | 0.75 | 0.23 | 0.82 | 0.19 |
|  | actin - v1 | 0.80 | 0.16 | 0.80 | 0.18 |
|  | microtubule - v2 | 0.19 | 0.12 | 0.17 | 0.01 |
|  | actin - v2 | 0.30 | 0.19 | 0.18 | 0.01 |

**Table S2. Comparison of average speeds ⟨$v$⟩ and corresponding standard deviations $\sigma$ measured with PIV and DDM.** Average speed ⟨$v$⟩ and corresponding standard deviation $\sigma$ for actin and microtubule channels of the videos shown in part A of Movies S1-S3, measured by fitting: (left) PIV speed distributions to one (purple, orange) or two (magenta) Schulz distributions, or (right) DDM $D(q, \Delta t)$ curves to functions that use one (purple, orange) or two (magenta) Schulz speed distributions. Note that all speeds are statistically indistinguishable between the two measurement techniques.

**Movies:**

**Movie S1: Sample time-series of active actin-microtubule composite exhibiting *Slow* dynamics. (A)** The example time-series used to demonstrate *Slow* dynamics of actin filaments (green) and microtubules (red) in Figs 2A, 3 and 4A, and (**B-D**) three additional time-series showing *Slow* dynamics.

**Movie S2: Sample time-series of active actin-microtubule composite exhibiting *Fast* dynamics. (A)** The example time-series used to demonstrate *Fast* dynamics of actin filaments (green) and microtubules (red) in Figs 2A, 3 and 4A, and (**B-D**) three additional time-series showing *Fast* dynamics.

**Movie S3: Sample time-series of active actin-microtubule composite exhibiting *Multimode* dynamics.** **(A)** The example time-series used to demonstrate *Multimode* dynamics of actin filaments (green) and microtubules (red) in Figs 2A, 3 and 4A, and (**B-D**) three additional time-series showing *Multimode* dynamics.

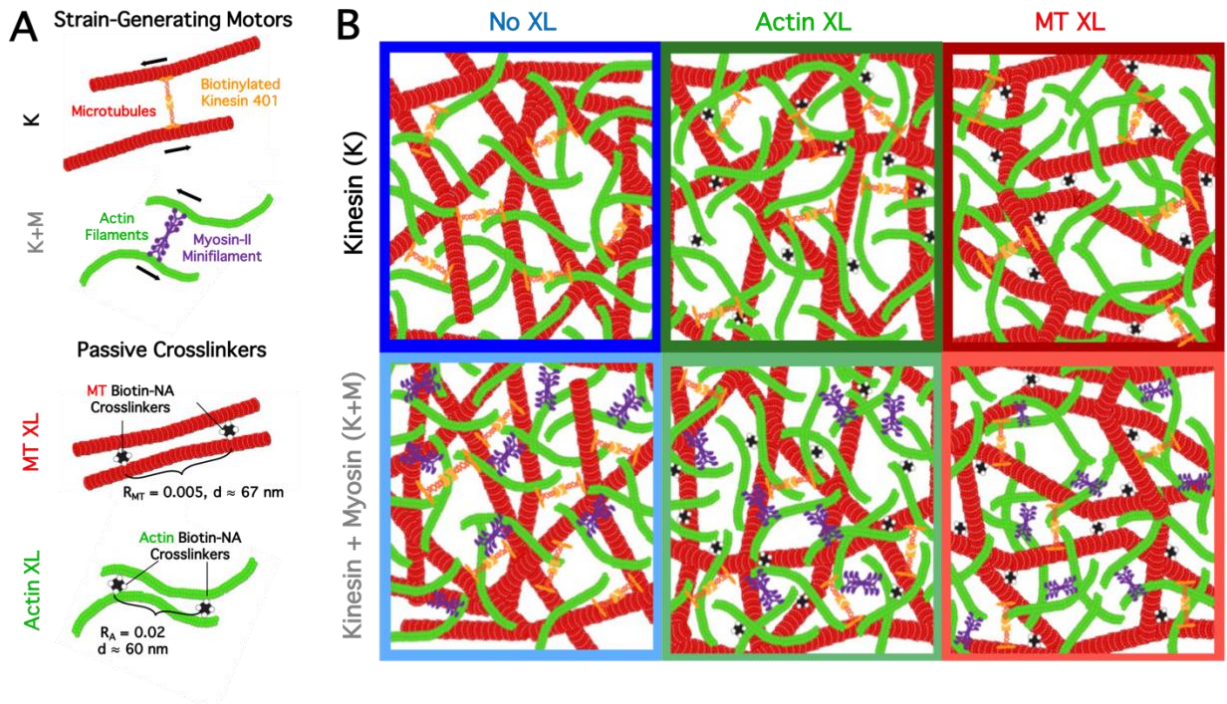

**Figure S1: Cartoon of phase space of different composite formulations.** **(A)** We co-polymerize actin monomers (2.32 µM) and tubulin dimers (3.48 µM) to form co-entangled composite networks of actin filaments (green) and microtubules (red). Static crosslinking is achieved using NeutrAvidin to link biotinylated actin filaments (Actin XL) or microtubules (MT XL). The crosslinker to protein molar ratio $R$ is fixed at $R = 0.02$ for actin and $R = 0.005$ for microtubules to achieve similar distances $d$ between crosslinks along the filaments. We incorporate kinesin clusters (orange) and myosin-II minifilaments (purple) as strain-generating motors to drive the composites out of steady-state. **B.** Cartoon of composite formulation space. We incorporate 0.35 µM kinesin (K) into composites with no static crosslinkers (No XL, dark blue box), actin-actin crosslinks (Actin XL, dark green box) and microtubule-microtubule crosslinks (MT XL, dark red box). For each kinesin-driven composite, we also examine the effect of adding 0.47 µM myosin (K+M) into composites with no static crosslinkers (No XL, light blue box), actin-actin crosslinks (Actin XL, light green box) and microtubule-microtubule crosslinks (MT XL, light red box).

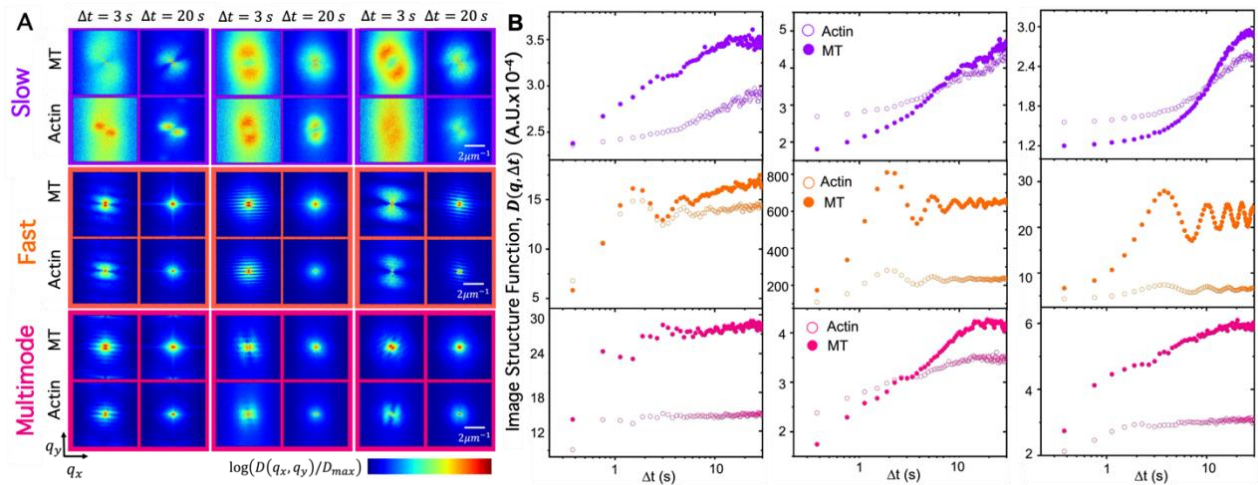

**Figure S2: Two-dimensional and azimuthally-averaged DDM image structure functions for 9 additional time-series with *Slow*, *Fast*, and *Multimode* behavior. A.** Two-dimensional image structure functions $D(q_x, q_y, \Delta t)$ computed for $\Delta t = 3$ s and $\Delta t = 20$ s for three representative time-series that display *Slow* (top rows, purple), *Fast* (middle rows, orange), and *Multimode* (bottom rows, magenta) characteristics. Colorscale is normalized separately for each image, and indicates the normalized value of each image structure function $[D(q_x, q_y, \Delta t)/D_{max}]$, with low (blue) and high (red) values indicative of lower or higher correlations. **B.** Azimuthally-averaged image structure functions $D(q, \Delta t)$ versus lag time $\Delta t$ computed from $D(q_x, q_y, \Delta t)$ functions shown in (A) for microtubule (closed symbols) and actin (open symbols) channels evaluated at $q = 1.33$ μm$^{-1}$.

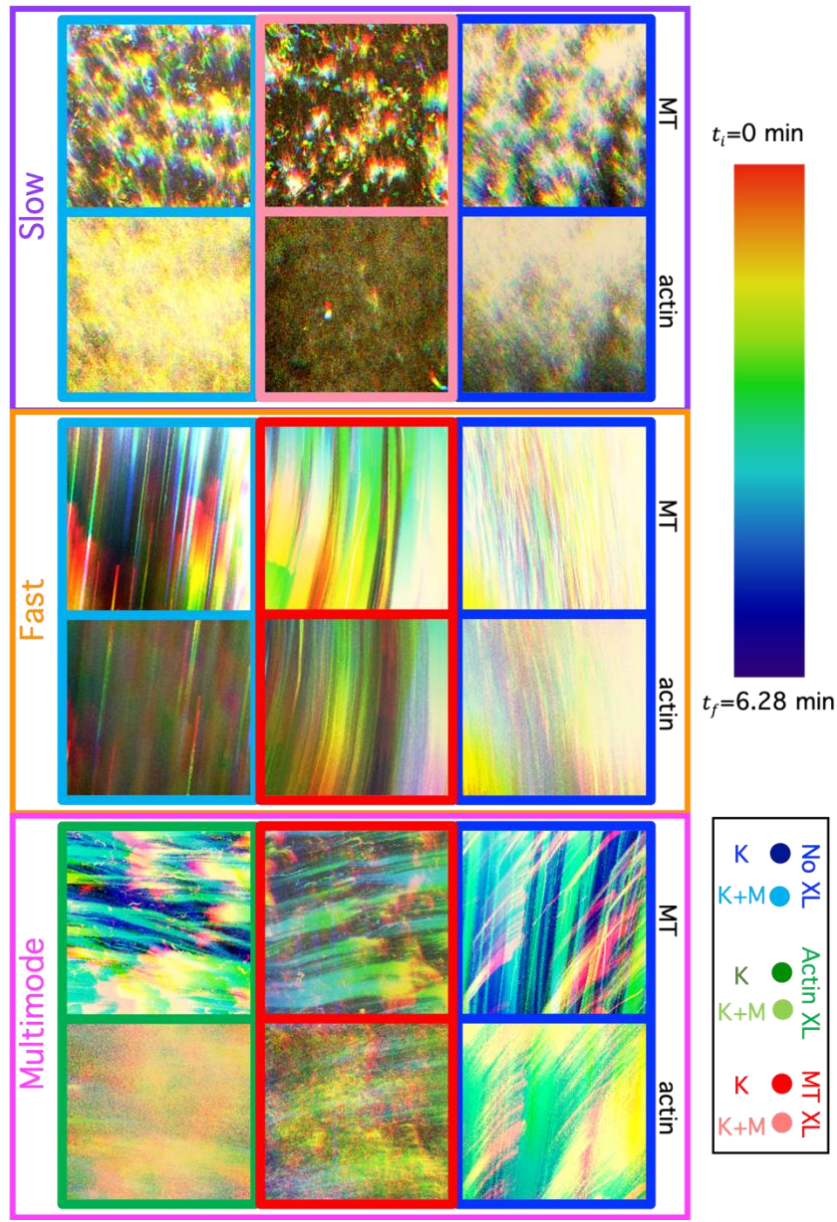

**Figure S3: Temporal color maps for 9 additional time-series with *Slow*, *Fast*, and *Multimode* dynamics.** Temporal color maps generated from the microtubule (top rows) and actin (bottom rows) channels of the nine different time-series analyzed in Fig S2, divided into *Slow* (top), *Fast* (middle), and *Multimode* (bottom) classes based on distinct $D(q,\Delta t)$ features shown in Fig 2A. Temporal color maps which colorize the features in each frame according to the time $t$ the frame is captured during the video, as indicated by the colorscale ($t_i = 0$ min (red) to $t_f = 6.28$ min (purple)), depict the motion of the composites. The color outlining each map denotes the composite formulation according to the legend. Each 256 × 256 square-pixel image is 213 μm × 213 μm. The videos from which maps are generated are parts B-D of Movies S1-S3.

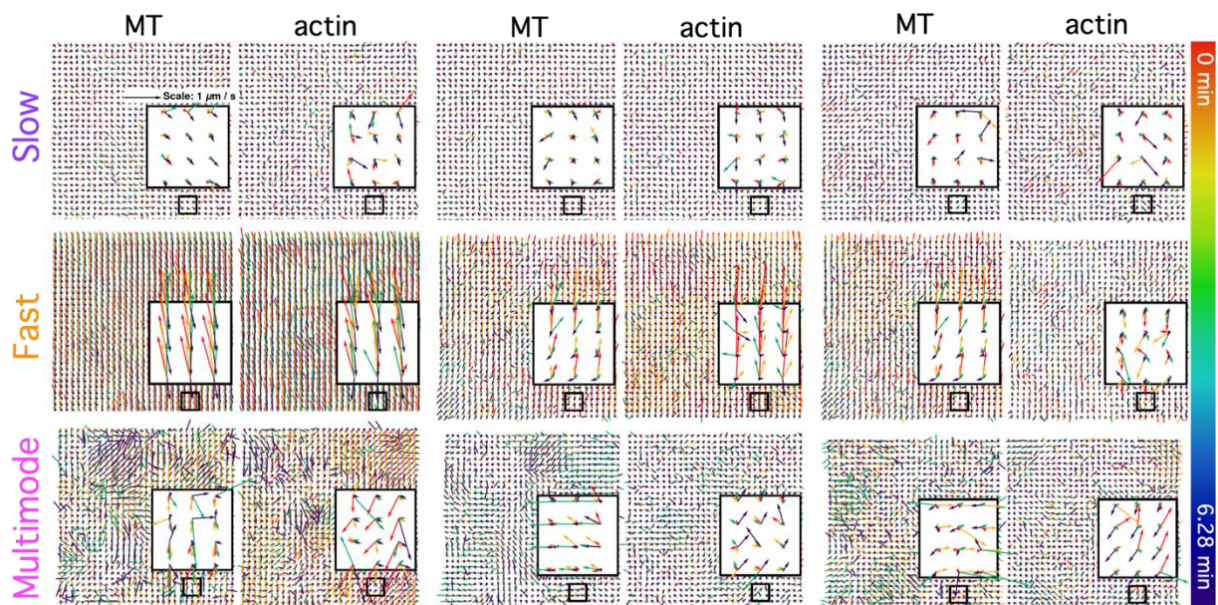

**Figure S4: PIV vector fields for 9 additional time-series with *Slow*, *Fast*, and *Multimode* dynamics.** PIV velocity vector fields for the microtubule (left) and actin (right) channels of parts B-D of Movies S1-S3 display *Slow* (top, purple, Movie S1), *Fast* (middle, orange, Movie S2), and *Multimode* (bottom, magenta, Movie S3) characteristics. Each arrow represents the average velocity vector for an $8 \times 8$ square-pixel region for $t = 0$ s (red), 125 s (yellow), 251 s (green) and 377 s (purple) as shown by the time-color scale. All vector fields are 213 µm × 213 µm and insets are zoom-ins of 25 µm × 25 µm square regions as indicated in the top-left field.

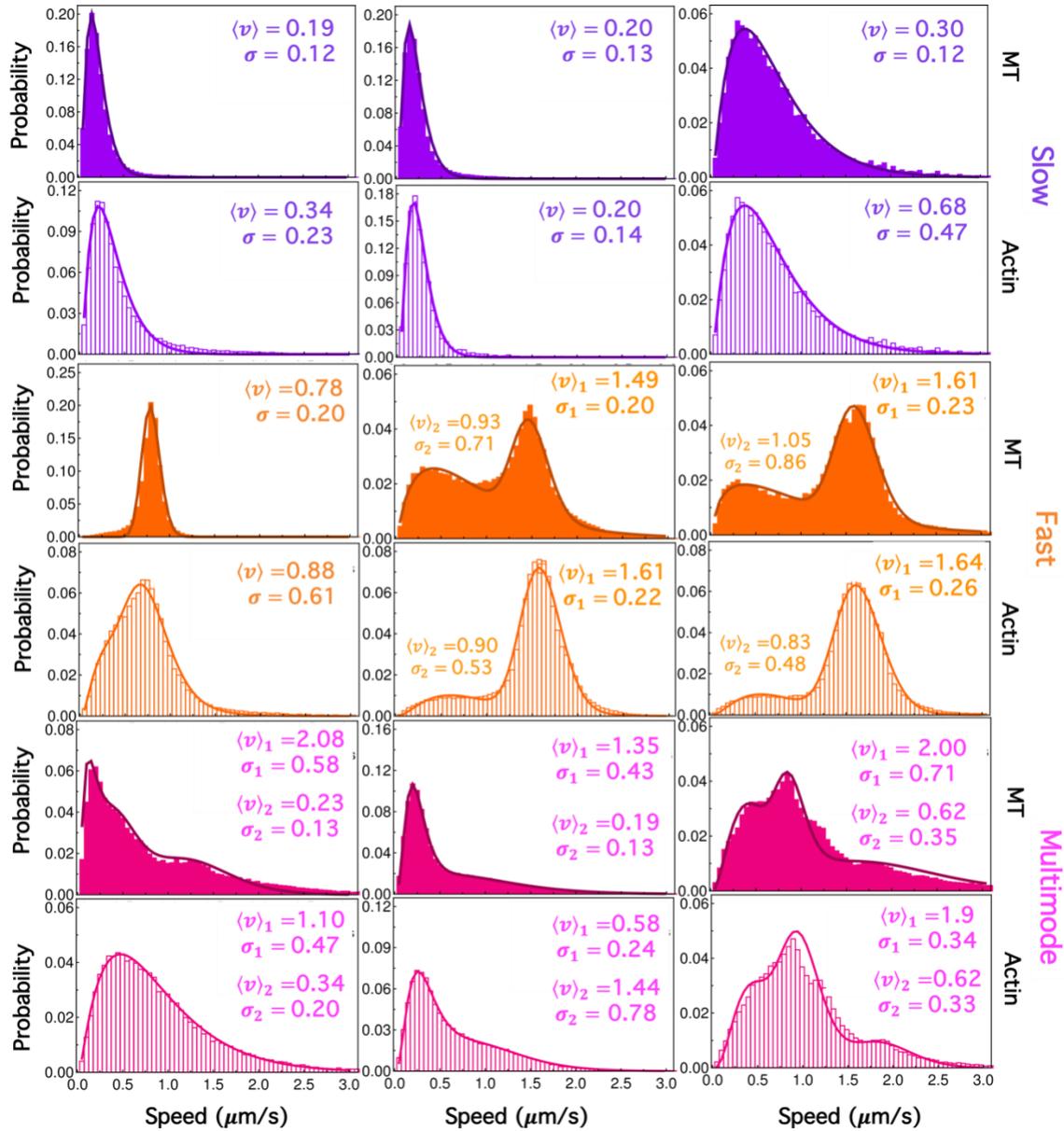

**Figure S5: Fits of 9 additional *Slow*, *Fast*, and *Multimode* speed distributions to Schulz functions.** Probability distributions of speeds determined from PIV for microtubules (filled) and actin (open) for the 9 *Slow* (top), *Fast* (middle) and *Multimode* (bottom) time-series analyzed in Fig S4 (B-D of Movies S1-S3). Dashed lines are fits to one or two Schulz distributions: $P(v) = \frac{v^z}{Z!}\left(\frac{Z+1}{\bar{v}}\right)^{Z+1} \exp\left[-\frac{v(Z+1)}{\bar{v}}\right]$ where $Z = \left(\frac{\bar{v}}{\sigma}\right)^2 - 1$ and $\bar{v}$ and $\sigma$ are the average and standard deviation of the speed distribution. $\bar{v}$ and $\sigma$ determined from each fit are listed in units of µm/s. *Multimode* distributions are best fit to a sum of two distributions with different $\bar{v}$ and $\sigma$ values (denoted by subscripts 1 and 2). Some *Fast* distributions are also better fit to a sum of two Schulz distributions, but the second distribution is weighted significantly less than the first distribution.

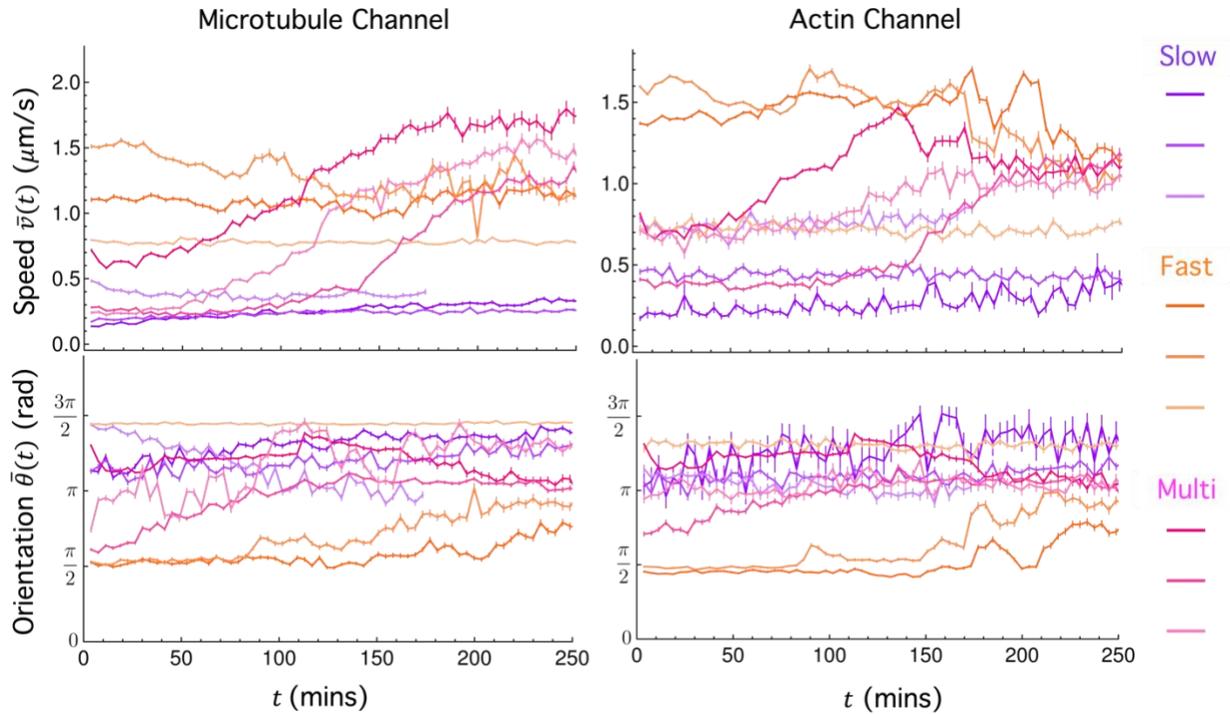

**Figure S6: Short-time course of average filament speeds and orientations for 9 additional time-series with *Slow*, *Fast*, and *Multimode* dynamics. (Top)** Average speed $\bar{v}(t)$ versus time $t$ measured via PIV for the MT (left) and actin (right) channels of the 9 representative *Slow* (purple), *Fast* (orange) and *Multimode* (magenta) videos analyzed in Fig S4 (B-D of Movies S1-S3). $\bar{v}(t)$ for each time $t$ is an average over all vector magnitudes in the PIV flow field associated with time $t$. **(Bottom)** Average velocity orientations $\bar{\theta}(t)$ versus $t$ computed from the same vector fields following the same method as for $\bar{v}(t)$.
16

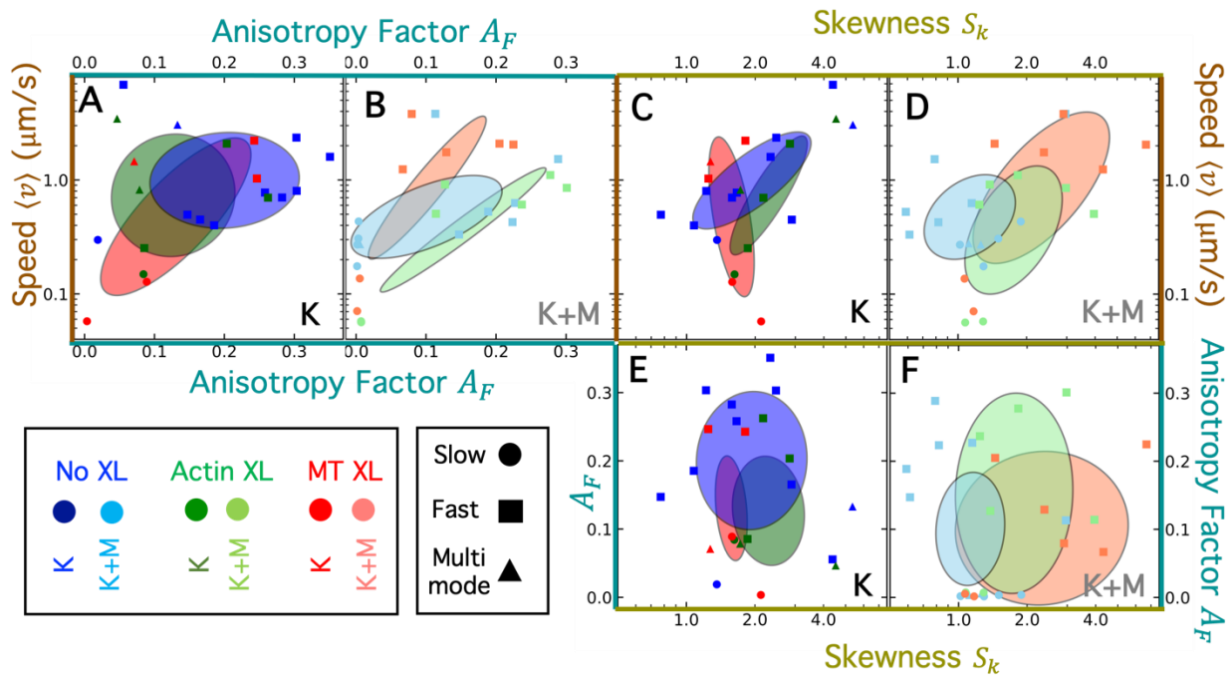

**Figure S7: Stacked 3-dimensional confidence ellipse plots show the relationships between average speed $\langle v \rangle$ (brown axes), anisotropy factor $|A_F|$ (teal axes), and skewness $S_K$ (gold axes) for different composite formulations.** Data points correspond to the 106 data points plotted in Fig 5, with colors and symbols indicating the composite formulation and dynamic class, respectively, according to the legends. The ellipses enclose one standard deviation around the mean. Panels with darker shaded (A,C,E) and lighter shaded (B,D,F) ellipses display data for composites with kinesin (K) and both kinesin and myosin (K+M), respectively.

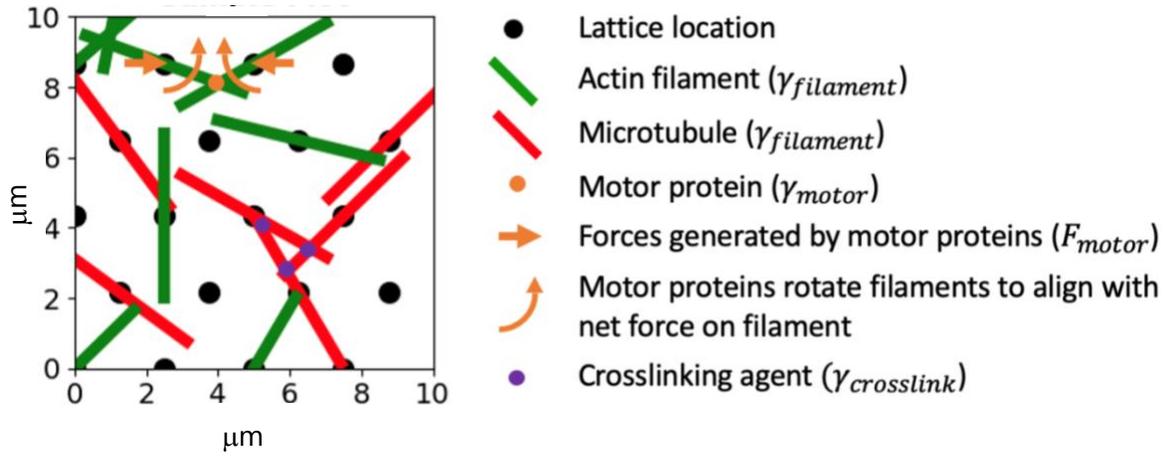

**Figure S8: Sample plot showing simulation mechanics.** The actin filaments and microtubules exist on a lattice of grid points. There is a drag ($\gamma_{filament}$) associated with their movement. Motor proteins exert forces which drive movement of the filaments ($F_{motor}$) but also exert drag ($\gamma_{motor}$). Crosslinking agents inhibit movement by exerting drag on their respective filaments ($\gamma_{crosslink}$).

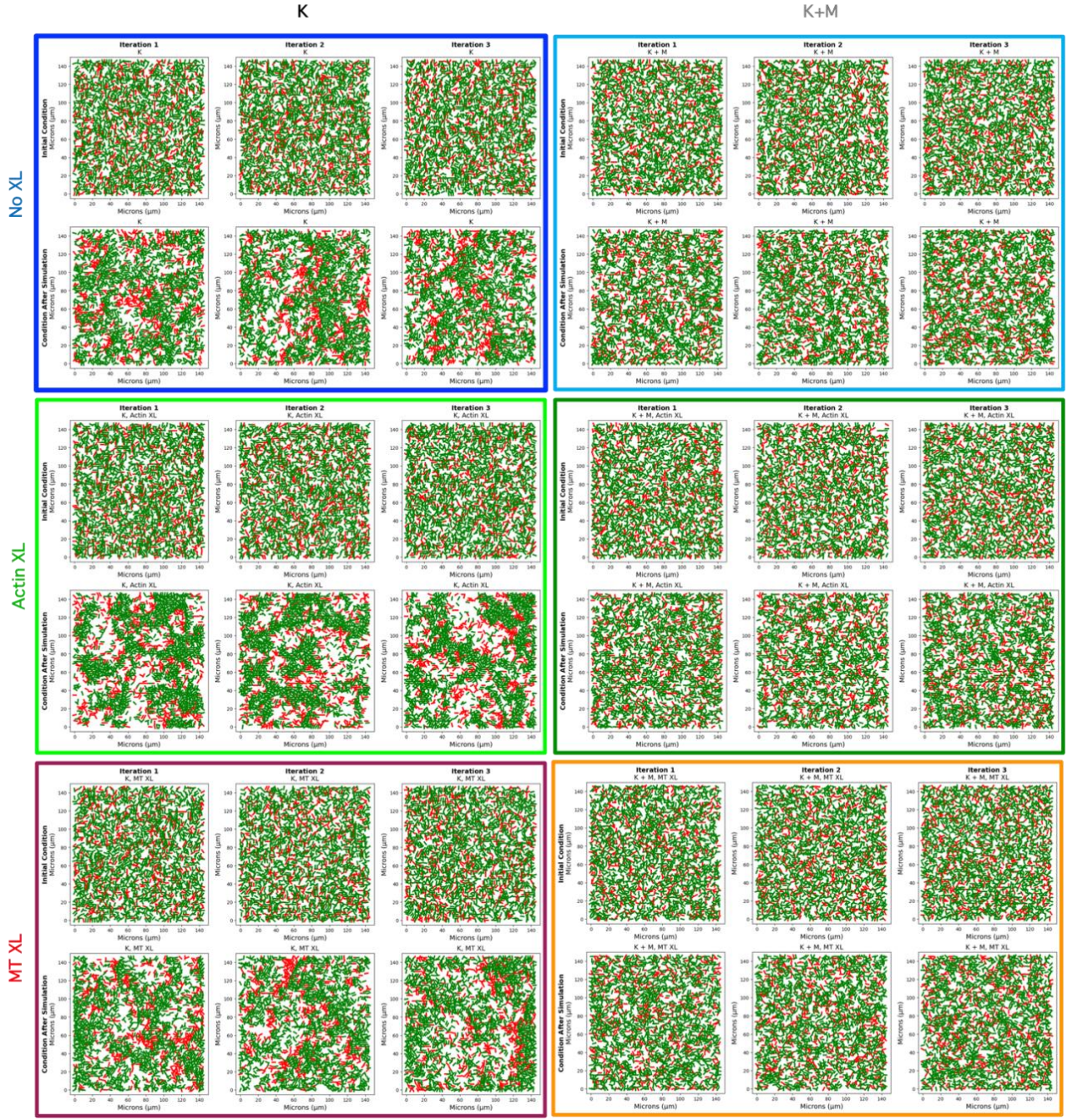

**Figure S9: Simulation snapshots for three independent trials for each composite formulation.** For each of the six composite formulations we investigate (indicated by the color-coded borders and labels) we simulate three independent iterations of the model. Correlation analysis data for each formulation (see SI Methods and Fig 6) are the corresponding average and standard error across the three trials. Color-coded borders enclose each formulation with lighter (right) and darker (left) shades denoting composites with and without myosin, respectively. For each formulation, the 6 images correspond to the initial (top row) and final (bottom row) states of the three independent runs (columns 1-3). All images show actin (green) and microtubules (red) comprising a 150 μm x 150 μm grid.

**Supplemental Information References**